\documentclass[10pt]{scrartcl}

\usepackage{garamondlibre,lmodern}
\usepackage{preprint}

\DeclareMathOperator{\diag}{diag}
\DeclareMathOperator*{\argmin}{arg\,min}
\DeclareMathOperator{\sgn}{sgn}
\DeclareMathOperator{\realpart}{real}
\DeclareMathOperator{\imagpart}{imag}

\def\placeholder{\textcolor{gray}{\ensuremath{\bullet}}}
\def\sm{\hphantom{-}}

\def\conv{\ast}
\def\xcorr{\star}

\definecolor{successful}{HTML}{335083}
\definecolor{unsuccessful}{HTML}{75211F}
\definecolor{failed}{HTML}{A13C0C}
\definecolor{rd}{HTML}{87160E}
\definecolor{gr}{HTML}{0E8740}
\definecolor{bl}{HTML}{003b7a}
\definecolor{pr}{HTML}{940774}

\usepackage[backend=bibtex,style=numeric-comp,sorting=none,backref=true]{biblatex}

\begin{document}\pagenumbering{arabic}


\addaffiliation[ESSEN]{Institute of Structural Analysis, University of Duisburg--Essen, 45141 Essen, Germany}%
\addaffiliation[PADERBORN]{Measurement Engineering Group, Paderborn University, 33098 Paderborn, Germany}%
\addaffiliation[MAGDEBURG]{Institute of Materials, Technologies and Mechanics, Otto von Guericke University Magdeburg, 39106 Magdeburg, Germany}%

\begin{header}%
	\begin{title}
		A modified Levenberg-Marquardt method for estimating the elastic material parameters of polymer waveguides using residuals between autocorrelated frequency responses
	\end{title}
	\begin{authors}
		\item Dominik~Itner\ESSEN\corresponding{dominik.itner@uni-due.de}
		\item Dmitrij~Dreiling\PADERBORN
		\item Hauke~Gravenkamp\MAGDEBURG
		\item Bernd~Henning\PADERBORN
		\item Carolin~Birk\ESSEN
	\end{authors}
	\begin{abstract}
		In this contribution, we address the estimation of the frequency-dependent elastic parameters of polymers in the ultrasound range, which is formulated as an inverse problem.
	This inverse problem is implemented as a nonlinear regression-type optimization problem, in which the simulation signals are fitted to the measurement signals.
	These signals consist of displacement responses in waveguides, focusing on hollow cylindrical geometries to enhance the simulation efficiency.
	To accelerate the optimization and reduce the number of model evaluations and wait times, we propose two novel methods.
	First, we introduce an adaptation of the Levenberg-Marquardt method derived from a geometrical interpretation of the least-squares optimization problem.
	Second, we introduce an improved objective function based on the autocorrelated envelopes of the measurement and simulation signals.
	Given that this study primarily relies on simulation data to quantify optimization convergence, we aggregate the expected ranges of realistic material parameters and derive their distributions to ensure the reproducibility of optimizations with proper measurements.
	We demonstrate the effectiveness of our objective function modification and step adaptation for various materials with isotropic material symmetry by comparing them with a state-of-the-art optimization method.
	In all cases, our method reduces the total number of model evaluations, thereby shortening the time to identify the material parameters.
	\end{abstract}
	\begin{keywords}
		\item material parameter estimation
		\item waveguide
		\item nonlinear optimization
		\item inverse problem
		\item least squares
	\end{keywords}
\end{header}


\section{Introduction}

The estimation of material parameters, specifically elastic and damping coefficients, for the application of materials in the ultrasound regime constitutes an important area of research with applications in both scientific and engineering domains.
In particular, the rapid and accurate identification of these parameters is paramount in industrial settings, such as quality control.
Although parameter estimation in the static and quasi-static regimes is well established, the identification of parameters in the ultrasound regime for high-performance applications is a relatively recent research question~\cite{Rautenberg2015,Bause2015,Bause2016}.
Inverse problems of this kind are solved by comparing some measurable quantities $\hat{\arr{y}}$ (such as displacements or strains) with the corresponding simulated quantities $\arr{f}(\arr{x})$ and minimizing the sum of squares of residuals
\begin{equation}\label{eq:objective}
	J(\arr{x}) = \norm{\hat{\arr{y}} - \arr{f}(\arr{x})}_2^2
\end{equation}
to fit the measurement and simulation by varying the parameters of interest~$\arr{x}$.

Several other attempts have been made based on different experimental setups~\cite{Sale2011,Johannesmann2022}, and direct inversion via neural networks~\cite{Held2024}.
These methods utilize the dispersion relations of broadband signals in the frequency-wavenumber domain based on measurements using laser beam excitations.
Conversely, the experimental setup we utilize is based on a comparison of transient signals with narrowband wave packets using piezoelectric transducers in a transmission setup~\cite{Bause2016}.
This setup has been extensively studied to determine the best possible experimental configuration and formulation of the associated simulation model~\cite{Dreiling2020,Dreiling2021,Dreiling2023,Dreiling2024}.

In this contribution, we primarily focus on the intricacies of the underlying optimization problem, including both the optimization method and modifications to the objective functions.
Some of the concepts proposed herein are generally applicable to other inverse problems.
However, we embed our proposition in the narrow context of the specific problem definition of material parameter estimation based on high-frequency signals.
Our primary goal is to provide a methodology in this setting that is both robust in finding the correct minimizer and fast by limiting the number of necessary optimization iterations, and hence, model evaluations.
This motivates the use of local optimization methods based on derivative information of a differentiable forward model.
However, to achieve this, the nonlinear objective function must be locally convex within the search space.

Typically, because the quantities to be compared are collected into vectors, the least-squares objective function in \Eqref{objective} is chosen to score the discrepancy between the measurement and simulation~\cite{Tam2020}.
The Gauss-Newton method and its extensions, such as the Levenberg-Marquardt method~\cite{Levenberg1944,Marquardt1963}, are particularly suited for this class of inverse problems.
The local convexity of the objective function is crucial for convergence to the minimizer for all local optimization methods, especially the Gauss-Newton method.
However, a direct comparison of the signals leads to highly non-convex optimization surfaces.
The Hilbert transform and its resulting envelope of signals have been shown to be remedies for this problem in the context of defect reconstruction in plates~\cite{Bulling2022}.
We explain how the envelope assists optimization in the context of material parameter estimation and suggest additional modifications in the frequency domain via the autocorrelation of the Fourier coefficients to further improve convergence of the optimization procedure.

Additionally, we propose a modification of the Levenberg-Marquardt method, which is purely derived from locally available information during the optimization process.
Many modifications and extensions of the Levenberg-Marquardt method have been proposed to select the damping or regularization parameter~$\lambda$.
These typically involve the objective function with ${\lambda \sim J(\arr{x})^q}$~\cite{Yamashita2001,Fan2005}, or more complex relationships involving the Jacobian of the forward model~\cite{Ma2007,Fan2009}, previous optimization steps~\cite{Fan2012,Fan2019,Li2023b}, or more intricate formulations~\cite{Karas2016,Amini2018,Li2023a,Han2024}.
We propose to consider the geometric interpretation of the nonlinear least-squares problem and demonstrate that the vector output of a forward model can be interpreted as points residing on a (relatively) low-dimensional manifold embedded in a high-dimensional ambient Euclidean space~\cite{Amari2016}.
This interpretation allows us to gain a strong intuition between the duality of the steepest descent method and the Gauss-Newton method and motivates our choice of $\lambda$.
This circumvents the  necessity for additional hyperparameters that would otherwise have to be tuned.

We evaluate our proposed methods by constructing a series of inverse problems involving three materials of interest: polyether-ether-ketone (PEEK), polycaprolactam or nylon 6 (PA6), and polypropylene (PP).
The simulation of virtual measurements provides access to the true material parameters, thereby supporting the validation of our optimization methods with ground truth data otherwise unavailable for measurements of real specimens.
We conduct a survey of the expected parameter ranges of the three materials to ensure the accuracy and reproducibility of these tests with actual measurements.
Using this data, we derive marginal distributions and draw samples from these to establish virtual references.
We present the objective function surface within the material ranges and demonstrate its convexification over the entire domain.
Subsequently, we optimize the individual materials, compare our method with the Broyden–Fletcher–Goldfarb–Shanno (BFGS) method, and demonstrate that our approach requires fewer model evaluations on average.

The remainder of this paper is organized as follows.
In \Secref{problem}, we introduce the experimental setup and corresponding forward model of wave propagation in the elastic domain, which forms the basis of the inverse problem.
In \Secref{optimization}, we motivate and describe our novel local optimization step scheme and the objective function modification based on signal transformations.
In \Secref{analysis}, we discuss the statistical analysis of the optimization behavior of a class of chosen materials and demonstrate the improved convergence behavior of the proposed methods based on virtual reference measurements.
Finally, in \Secref{conclusion}, we present the conclusion and final remarks, as well as an outlook for future research.

\section{Problem description: Experimental setup and simulation via forward model}
\label{sec:problem}

The broader context of our work is to estimate the elastic parameters of polymers at high frequencies by measuring the displacement response in waveguides.
As we are primarily concerned with the dynamic elastic behavior of homogeneous materials in the high-frequency regime, where the small-strain assumption holds due to small excitation energies, the model needs to satisfy the following linear equations of motion without body loads (Einstein notation implied)
\begin{equation}\label{eq:differential:equation}
	C_{ijkl} \partial_{il} u_k - \rho \partial_{tt} u_i = 0
	\punct,
\end{equation}
with the elastic tensor~$C_{ijkl}$, displacements~${u_i = u_i(r, \theta, z, t)}$, and cylindrical coordinates $i, j, k, l \in \{r, \theta, z\}$.
For the assessment of the proposed optimization method and objective function transformation, we assume isotropic material symmetry for the virtual measurements with the corresponding elasticity matrix with Young's modulus~$E$ and Poisson's ratio~$\nu$
\begin{equation}
	\arr{C}\subrm{iso}\inverse
	=
	\begin{bmatrix}
		\hphantom{-}\frac{  1}{E} &           - \frac{\nu}{E} &           - \frac{\nu}{E} & & & \\[0.5mm]
		          - \frac{\nu}{E} & \hphantom{-}\frac{  1}{E} &           - \frac{\nu}{E} & & & \\[0.5mm]
		          - \frac{\nu}{E} &           - \frac{\nu}{E} & \hphantom{-}\frac{  1}{E} & & & \\
		& & & \frac{2(1 + \nu)}{E} &                      &                      \\
		& & &                      & \frac{2(1 + \nu)}{E} &                      \\
		& & &                      &                      & \frac{2(1 + \nu)}{E}
	\end{bmatrix}
	\punct.
\end{equation}

\begin{figure}[H]%
	\begin{minipage}[b]{0.48\linewidth}%
		\sameheight{0.99\linewidth}[figures/setup/]{measurement,simulation}%
		\centering%
		\subcaptionbox{\label{subfig:setup:measurement}}%
		{%
			\includegraphics[height=\samesubheight]{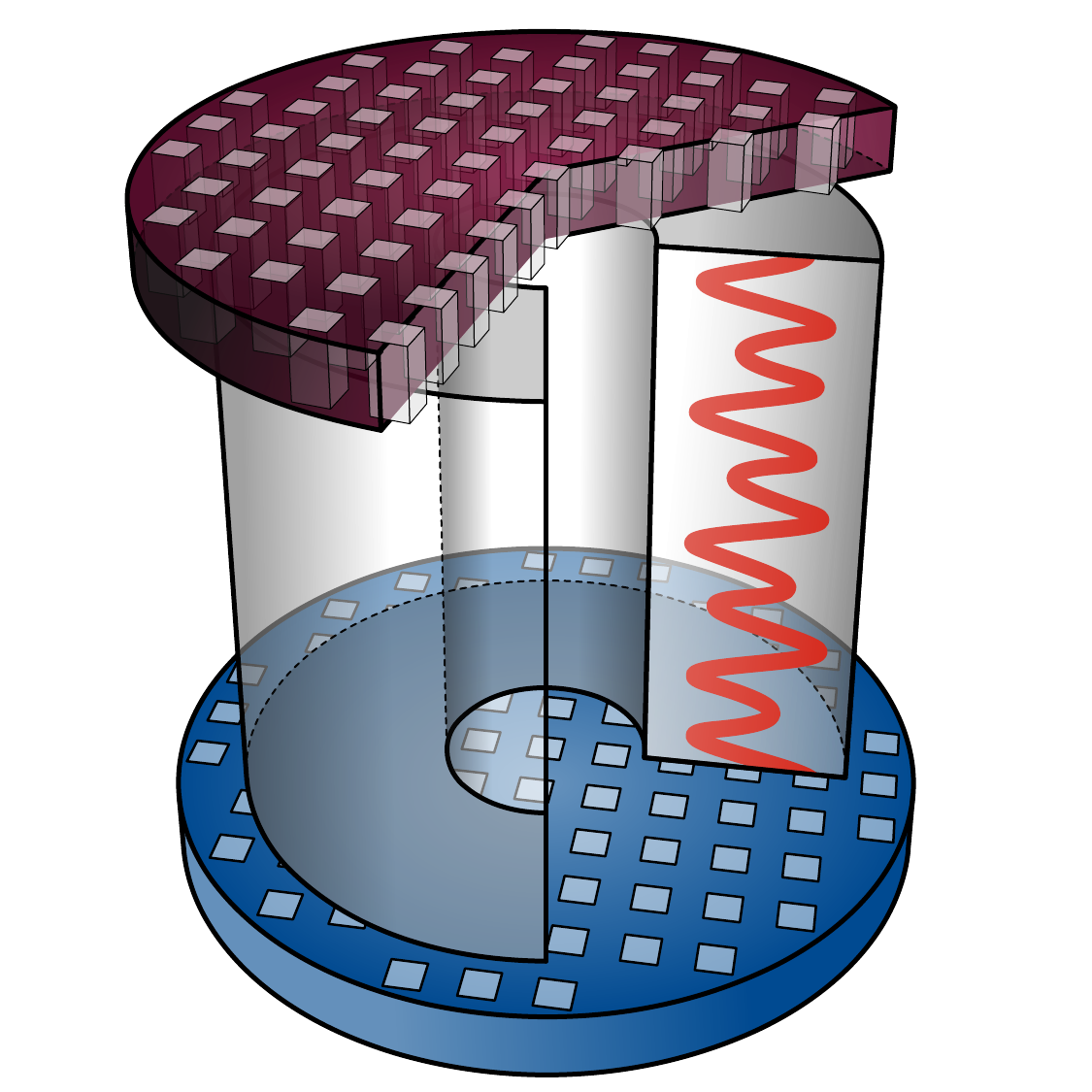}%
			\vspace*{-5mm}%
		}%
		\hfill%
		\subcaptionbox{\label{subfig:setup:simulation}}%
		{%
			\includegraphics[height=\samesubheight]{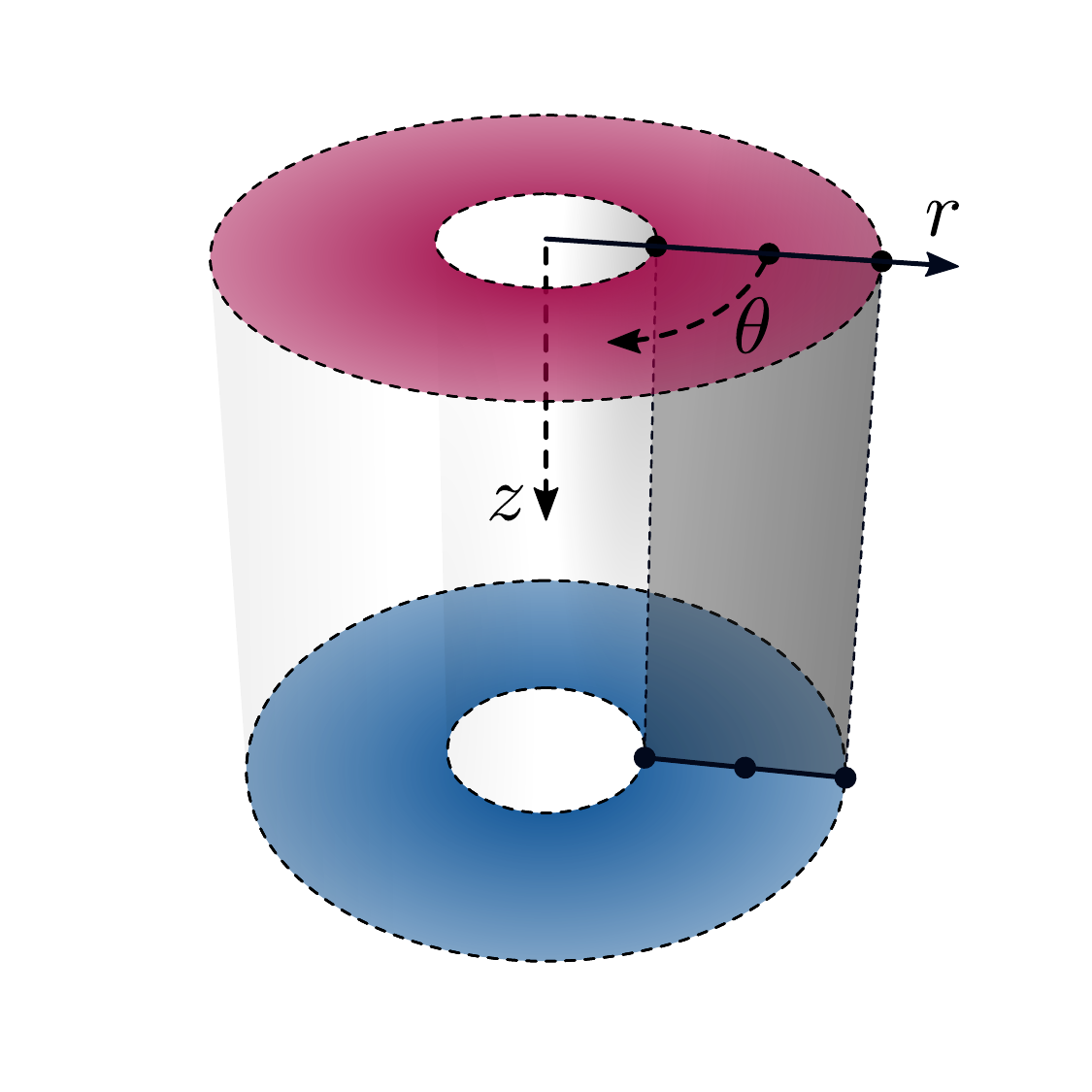}%
			\vspace*{-5mm}%
		}%
		\captionsetup{width=\linewidth}%
		\caption%
		{%
			Schematic of the measurement setup with cylindrical sample~\subfigref{setup}{measurement} and the accompanying representation of the simulated domain~\subfigref{setup}{simulation}.
			The FEM nodes are indicated as dots ($\bullet$) in the semi-discretized domain.\\
		}%
		\label{fig:setup}%
	\end{minipage}%
	\hfill%
	\begin{minipage}[b]{0.48\linewidth}
		\centering%
		\includegraphics[width=\linewidth]{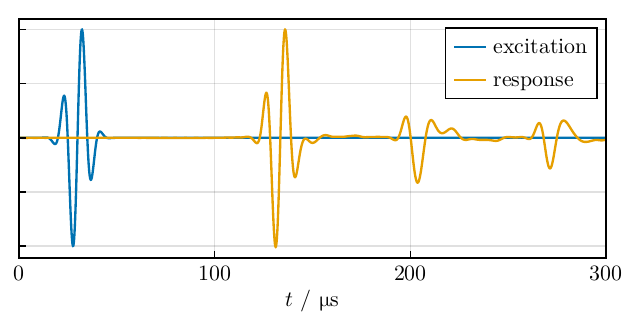}%
		\vspace*{-5mm}%
		\caption{%
			Normalized excitation wave packet and simulated response of the transmission setup.
			Secondary and tertiary wave packets in the response are of special interest, as they are responsible for the sensitivity to shear-related material parameters.
		}%
		\label{fig:example:signal}%
	\end{minipage}
\end{figure}

The experimental measurement is set up as follows~\cite{Bause2016}.
The sample has a hollow cylindrical geometry with a height and diameter of \SI{20}{\milli\meter} and \SI{19.08}{\milli\meter}, respectively.
Both the excitation and measurement sites are located at the annular faces of the cylinder.
Piezoelectric 1-3 composite-based ultrasound transducers function as a transmitter-receiver pair~\cite{Dreiling2021}.
An electrical signal is provided, and the piezoelectric composite converts it into a mechanical excitation (here, a wave packet) that propagates through the sample, which acts as a waveguide.
See~\Subfigref{setup}{measurement} for a schematic of the setup and \Figref{example:signal} for a simulated example of the excitation and response.
The mechanical wave is detected and measured at the opposite end of the cylinder via the receiver, which converts the mechanical displacements of the surface into a single electrical signal.
This conversion of electrical to mechanical signal, and vice versa, is approximated by a simple one-dimensional Mason model~\cite{Smith1991,Dreiling2021}.
A more elaborate three-dimensional simulation of mechanical wave propagation utilizes a semi-analytic formulation based on the scaled boundary finite element method (SBFEM)~\cite{Gravenkamp2015,Song2018,Gravenkamp2018,Itner2021Ultrasonics}.
In the present study, we assume a full-face excitation.

The SBFEM allows for an efficient simulation of guided wave propagation in cylindrical domains.
In this setting, only the radial direction must be discretized using the principles of the finite element method (FEM).
The displacement field in the circumferential direction is decomposed via a periodic Fourier transform, such that each associated spatial mode can be solved independently.
The wavefield dependence in the axial direction is described analytically via the SBFEM formulation.
This reduces the three-dimensional problem for all practical purposes to a one-dimensional problem along the radial direction, considering the typical sizes of system matrices, such as the stiffness and mass matrices.
The schematic in \Subfigref{setup}{simulation} demonstrates the reduction in the degrees of freedom in the domain.
A sufficient problem-dependent number of degrees of freedom is chosen for all simulations~\cite{Itner2021Ultrasonics}.
The dynamic forward problem in \Eqref{differential:equation} is then solved in the frequency domain, likewise for each temporal frequency.
This formulation yields a frequency-dependent linear relationship between the displacements and tractions of the typical form
\begin{equation}
	\arr{K}(\omega) \arr{u}(\omega) = \arr{p}(\omega)
	\punct,
\end{equation}
for nodal displacements $\arr{u}$ and tractions $\arr{p}$ along the opposing faces of the cylinder, and dynamic stiffness matrix $\arr{K}$ for a particular (temporal) angular frequency $\omega$.
The time dependence of the traction at the excitation site is given by
\begin{equation}\label{eq:excitation}
	p_z(t)
	=
	\sin(2\pi \, \bar{f} \, t) \, \euler^{-\frac{(t - \bar{t})^2}{2 \sigma^2}}
	\punct,
\end{equation}
where $\sigma$ is chosen to give a bandwidth-limited signal in the frequency domain with $b = 0.65\bar{f}$ dependent on the center frequency~$\bar{f}$
\begin{equation}
	\sigma = (\pi \, b)^{-1}
	\punct.
\end{equation}
In~\Figref{example:signal}, an exemplary (normalized) excitation wave and the corresponding response are shown for the transmission setup.
We choose $\bar{f} = \SI{1}{\giga\hertz}$ and $\bar{t} = \SI{3}{\micro\second}$ for all examples.
Details of the formulation and implementation can be reviewed in~\cite{Itner2021Ultrasonics}.

Given that the receiver is designed as a basic ultrasound transducer, its output is a singular electrical signal, resulting in the measurement being a scalar function of time.
However, the forward model yields the entire displacement field at the corresponding face.
Therefore, the measurement is not easily compared to the simulated response.
Evaluating the response at only a single point on the face would be too great a simplification of the measurement.
We assume that the conversion of mechanical displacements into an electric potential implicitly performs an area averaging of the displacement field.
After solving for the unknown degrees of freedom, the displacement field of the model can be reconstructed as follows
\begin{equation}\label{eq:displacementfield}
	\arr{u}
	=
	\sum_{m \in \mathbb{M}} \arr{N}(\eta) \hat{\arr{u}}_m(z, t) \, \euler^{\imag m \theta}
	\punct,\quad
	\arr{u}\transpose
	=
	\begin{bmatrix}
		u_r(r, \theta, z, t) & u_\theta(r, \theta, z, t) & u_z(r, \theta, z, t)
	\end{bmatrix}
	\punct,
\end{equation}
with a matrix of shape functions $\arr{N}$ w.r.t.\ the local coordinate $\eta \in [-1,+1]$ of a single finite element, nodal vector of Fourier coefficients $\hat{\arr{u}}_m(z, t)$, and circumferential harmonic wavenumber $m$.
Here, the set of harmonic indices $\mathbb{M}$ and associated Fourier coefficients are a prefiltered subset required for a convergent solution $\mathbb{M} \subset \integers$ of all harmonic integers of the discrete Fourier series of an arbitrarily shaped boundary condition~\cite{Itner2021Ultrasonics}.
Integration of the displacement field over the annular face at $z$ with
\begin{equation}\label{eq:integral:definition}
	\bar{\arr{u}}(\vartheta, z, t)
	=
	\int_0^\vartheta \int_{-1}^{+1} \sum_{m \in \mathbb{M}} \arr{N}(\eta) \, \hat{\arr{u}}_m(z, t) \euler^{\imag m \theta} \, \bar{\arr{J}}(\eta) \dd{\eta} \dd{\theta}
	\punct,\quad
	\bar{\arr{J}}(\eta)
	=
	\frac{r}{2} \ell + \frac{1 + \eta}{4} \ell^2
	\punct,
\end{equation}
allows to compute the area average~$y$
\begin{equation}\label{eq:areaaverage}
	y(\theta_1, \theta_2, z, t)
	=
	2 \frac{\bar{u}_z(\theta_2, z, t) - \bar{u}_z(\theta_1, z, t)}{(\theta_2 - \theta_1) (r\subrm{o}^2 - r\subrm{i}^2)}
	\punct,
\end{equation}
where $\bar{u}_z$ is the $z$-component of $\bar{\arr{u}}$.
In our particular case for $\theta_1 = 0$ and $\theta_2 = 2\pi$ and $\mathbb{M} = \{0\}$ \Eqref{areaaverage} simplifies to
\begin{equation}
	y(z, t)
	=
	\frac{\bar{u}_z(z, t)}{\pi(r\subrm{o}^2 - r\subrm{i}^2)}
	\punct,\quad
	\bar{\arr{u}}(z, t)
	=
	\int_{-1}^{+1} \arr{N}(\eta) \, \hat{\arr{u}}_0(z, t) \, \bar{\arr{J}}(\eta) \dd{\eta}
	\punct,
\end{equation}
which can be computed with great precision via numerical integration of the shape functions.
Finally, the time-dependent area~average~$y(z,t)$ and measurement~$\hat{y}(t)$ can be used to formulate an adequate objective function based on the sum of squares (where $z$ is implied to be at the location of the measurement)
\begin{equation}\label{eq:optimization:objective}
	\hat{\arr{x}}
	=
	\argmin_{\arr{x}}
	\sum_{t_i} \tfrac{1}{2} \big( \hat{y}(t_i) - y(t_i) \big)^2
	\punct,
\end{equation}
to find the minimizer $\hat{\arr{x}}$.
Here, $y(t_i)$ depends implicitly on the parameters $\arr{x}$ to reduce notational overload.

\section{Optimization method and objective function}
\label{sec:optimization}

The optimization objective for the material parameter estimation in \Eqref{optimization:objective} is a nonlinear, highly non-convex optimization problem.
The objective is stated as a least-squares problem, due to the nature of the compared quantities.
An inverse problem is typically solved by minimizing the rather straightforward objective function $J(\arr{x})$ and the optimization is discussed primarily in the context of convergence rates under the assumption of convexity.
In this context, the least-squares formulation is often reduced to a mathematical abstraction intended to quantify differences between states in some ostensibly meaningful way, although the precise interpretation of "meaningful" is rarely examined further.

However, we claim that a strong geometrical understanding is hidden behind the otherwise abstract notion of the nonlinear least-squares problem~\cite{Amari2016,Tarantola2005}.
The geometric interpretation is inherent to the problem because the goal is to minimize the length of the residual vector.
This hides a strong intuition about the problem at hand and how to possibly improve the optimization behavior without the need for hyperparameters that require further fine-tuning.

In the following, we briefly develop the geometrical intuition required to motivate our proposed modification.
We show in passing how the classical Gauss-Newton method is obtained trivially in the context of this formulation, and then understand the standard gradient descent method as a dual to the Gauss-Newton method and their geometrical dual relationship as orthogonal projections.
Finally, we demonstrate why the gradient descent step is geometrically questionable~\cite{Strutz2010,Tarantola2005}, propose strategies to correct this, and introduce a novel automatic local step-size adaption for the Levenberg-Marquardt method.

\subsection{The model manifold}
\label{sec:optimization:modelmanifold}

The forward model used in this study is based on the equations of motion governing linear elasticity.
In this sense, the model is linear because of the small-strain formulation.
In contrast, the influence of the material parameters on the solution is highly nonlinear due to variations in the amplitudes of the response signals and arrival times of the primary and subsequent wave packets, as well as phase shifts in the carrier frequency.
This nonlinearity leads to the necessity of iterative optimization methods in the first place but is also responsible for the difficulty of developing robust optimization methods.
The question may then arise: What does nonlinearity mean in this context specifically, and how can it be visualized to gain intuition?
In principle, nonlinearity exists in any system if non-zero second-order derivatives with respect to the quantities of interest exist.
However, the intuitive geometrical interpretation of the second-order derivative is simply \emph{curvature}.
Consequently, visualization of this curvature, i.e., visualization of the simulation model as the set of its solutions as its own geometrical entity, the model manifold, is possible.

First, the model manifold is defined.
All measurements and simulation responses exist for discrete time steps $t_i$.
We now explicitly introduce the dependency on the parameters to define the model manifold.
The previously described variables $y(t)$ and $\hat{y}(t)$ can be expressed as vector quantities
\begin{align}\label{eq:model:vectors}
	\hat{\arr{y}}
	&=
	\begin{bmatrix}
		\hat{y}(t_1) \\
		\vdots \\
		\hat{y}(t_n)
	\end{bmatrix} \in \reals^n
	\punct,
	&
	\arr{f}(\arr{x})
	&=
	\begin{bmatrix}
		y(t_1; x_1, \dots, x_m) \\
		\vdots \\
		y(t_n; x_1, \dots, x_m)
	\end{bmatrix} \in \reals^n
	\punct,
	&
	\arr{x}
	&=
	\begin{bmatrix}
		x_1 \\
		\vdots \\
		x_m
	\end{bmatrix} \in \reals^m
	\punct.
\end{align}
We denote by $\hat{\arr{y}}$ the reference vector, $\arr{f}(\arr{x})$ the forward model (which is a vector-valued vector function), and $\arr{x}$ the vector of parameters, which may be filled with $m$ independent variables of interest.
In our analysis, we consider isotropic material symmetry with
\begin{align}\label{eq:parameters}
	\arr{x}\subrm{iso}
	&=
	\begin{bmatrix}
		E & \nu
	\end{bmatrix}\transpose
	\in \reals^2
	\punct.
\end{align}
For conciseness and generality, the subscript of $\arr{x}$ is neglected.
The model manifold can then be defined as the set of all points $\arr{f}(\arr{x}) \in \reals^n$ in the space of all practically feasible and infeasible measurement signals that belong to parameters $\arr{x} \in \reals^m$ for our model
\begin{equation}\label{eq:manifold:definition}
	\mathcal{M} = \{ \arr{f}(\arr{x}) \in \reals^n \mid \arr{x} \in \reals^m \}
	\punct.
\end{equation}
It is a curved hypersurface embedded in an ambient signal space in $\reals^n$ with a curvilinear coordinate system spanned by the parameters, and consists of the set of all responses the forward model can produce, given a consistent description of the domain's geometry and boundary conditions\footnote{Here, domain refers to the spatial domain for which the equations of motion of linear elasticity are solved.}.

\begin{figure}[H]%
	\sameheight{0.75\linewidth}[figures/manifold/]{manifold-parameter,manifold-signal,manifold-envelope}%
	\definecolor{cora}{HTML}{0072B2}%
	\newsavebox{\cora}\savebox{\cora}{%
		\tikz[baseline=-0.5ex]{\draw[color=cora, line width=1.5pt, line cap=round] (0,0) -- (6pt,0);}%
	}%
	\definecolor{corb}{HTML}{D55E00}%
	\newsavebox{\corb}\savebox{\corb}{%
		\tikz[baseline=-0.5ex]{\draw[color=corb, line width=1.5pt, line cap=round] (0,0) -- (6pt,0);}%
	}%
	\centering%
	\hspace*{\stretch{0.1}}%
	\subcaptionbox{\label{subfig:model:manifold:parameters}}%
	{%
		\includegraphics[height=\samesubheight]{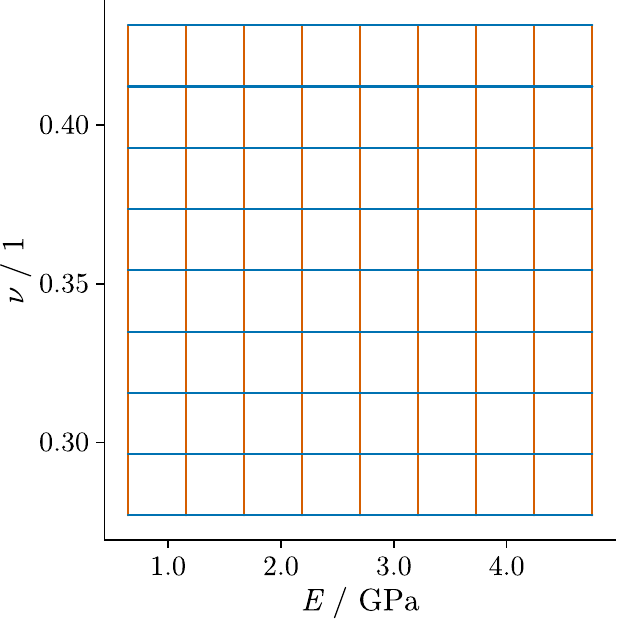}%
		\vspace*{-5mm}%
	}%
	\hspace*{\stretch{1}}%
	\subcaptionbox{\label{subfig:model:manifold:signals}}%
	{%
		\includegraphics[height=\samesubheight]{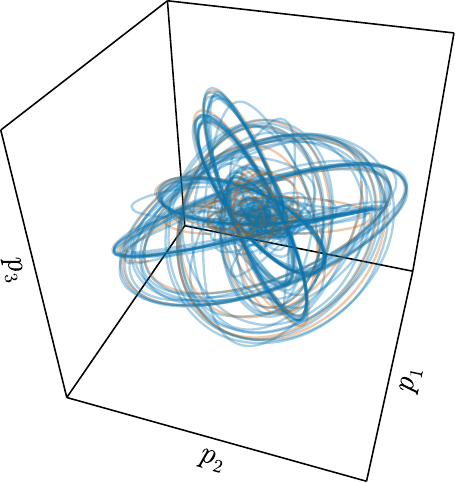}%
		\vspace*{-5mm}%
	}%
	\hspace*{\stretch{1}}%
	\subcaptionbox{\label{subfig:model:manifold:envelopes}}%
	{%
		\includegraphics[height=\samesubheight]{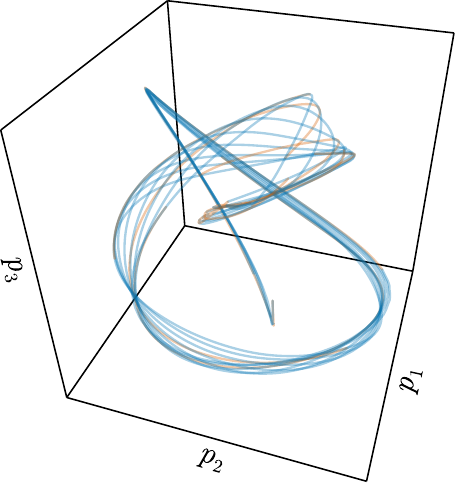}%
		\vspace*{-5mm}%
	}%
	\hspace*{\stretch{0.1}}%
	\captionsetup{width=\linewidth}
	\caption%
	{%
		Manifolds (and projections thereof) associated with the two-dimensional parameter space~\subfigref{model:manifold}{parameters} with $E$~(\usebox{\cora}) and $\nu$~(\usebox{\corb}), the signal space~\subfigref{model:manifold}{signals}, and the envelope space ~\subfigref{model:manifold}{envelopes}.
		The intersections in~\subfigref{model:manifold}{signals} are visual artifacts due to projection.
		The envelope as a nonlinear transformation reduces the complexity of the manifold, s.t.\ even in lower dimensions, no intersections occur~\subfigref{model:manifold}{envelopes}.
	}%
	\label{fig:model:manifold}
\end{figure}

This model manifold can be visualized by sampling its curvilinear coordinate system, as shown in \Figref{model:manifold}.
Note that for the model output $\arr{f}(\arr{x}) \in \reals^n$, usually we have $n \gg 3$.
Hence, any visualization requires a dimensionality reduction method to properly project the high-dimensional model manifold into, at most, a three-dimensional space.
To this end, we use the principal component analysis to linearly project the model output~$\arr{f}(\arr{x})$ onto the first three principal directions.
In \Subfigref{model:manifold}{parameters}, the parameter space itself is sampled along the constant coordinate lines of $E$ and $\nu$\footnote{The parameter space itself is a Euclidean space because we assume no correlations between the chosen parameters.}.
The associated parameter combinations are used as inputs into the forward model, and the resulting coordinate lines of the signals are presented in \Subfigref{model:manifold}{signals}.
This visualization demonstrates the difficulty of moving along the manifold surface to reach a particular minimizer for an arbitrary reference.
Additionally, we present the same manifold but transformed via the envelope function, i.e., the envelopes of the signals are generated and compared in \Subfigref{model:manifold}{envelopes}.
It is easy to see that moving along the surface of the envelope manifold leads to broader regions within which an initial estimate might lead to the correct minimizer.
In \Secref{objectivefunction:novel}, we delve deeper into this topic in order to clarify this observation.
Visualization becomes increasingly impractical for higher-dimensional manifolds, e.g., even the projection of a three-dimensional manifold into $\reals^3$ leads to false intersections that render any visualization uninterpretable.

By using the notation given in \Eqref{model:vectors}, the optimization objective in \Eqref{optimization:objective} is rephrased as vector quantities
\begin{equation}\label{eq:optimization:objective:vector}
	\hat{\arr{x}}
	=
	\argmin_{\arr{x}}{}
	\frac{1}{2} \norm{\arr{y} - \arr{f}(\arr{x})}^2
	\punct.
\end{equation}
Since we want to utilize first-order optimization methods based on gradient information of the form
\begin{equation}\label{eq:optimization:iteration}
	\arr{x}_{k+1} = \arr{x}_k + \Delta\arr{x}_k
\end{equation}
with increment $\Delta\arr{x}_k$ at iteration $k$, the differentiation of \Eqref{optimization:objective:vector} is paramount.
In steepest descent methods, the increment is typically chosen as the gradient $\Delta\arr{x}_k = -\alpha \, \nabla \frac{1}{2} \norm{\arr{y} - \arr{f}(\arr{x})}^2$ with step adaption~$\alpha$.
Differentiation w.r.t.\ any one parameter yields
\begin{equation}\label{eq:gradient:loss}
	\pdv{}{x_i} \frac{1}{2} \norm{\arr{y} - \arr{f}(\arr{x})}^2
	=
	-\big( \arr{y} - \arr{f}(\arr{x}) \big) \cdot \pdv{}{x_i}\arr{f}(\arr{x})
	\punct,
\end{equation}
which requires differentiation of the forward model.
Differentiation of the SBFEM with respect to the material parameters is possible, and differentiation of the SBFEM, in general, has been demonstrated in~\cite{Chowdhury2014}.
Therefore, we refrain from presenting the derivation here.
Collecting all derivatives into a single matrix results in the Jacobian matrix
\begin{equation}\label{eq:definition:jacobian}
	\arr{J}(\arr{x})
	=
	\begin{bmatrix}
		\pdv{}{x_1}\arr{f}(\arr{x}) & \dots & \pdv{}{x_m}\arr{f}(\arr{x})
	\end{bmatrix}
	\punct.
\end{equation}
Because the orders of magnitude of the individual material parameters may differ greatly (such as for $E$ and $\nu$), the column vectors of $\arr{J}$ may deviate by several orders of magnitude.
To alleviate this numerically problematic mismatch of scales, we rescale the Jacobian matrix using the local (current) material parameters, s.t.
\begin{equation}\label{eq:jacobian:rescale}
	\tilde{\arr{J}}(\arr{x}_k) = \arr{J}(\arr{x}_k) \diag(\arr{x}_k)
	\punct.
\end{equation}
Then any derived increment $\Delta\tilde{\arr{x}}_k$, which is discussed in \Secref{descentstep}, must similarly be rescaled via
\begin{equation}
	\Delta\arr{x}_k = \diag(\arr{x}_k) \Delta\tilde{\arr{x}}_k
	\punct.
\end{equation}
This rescaling can be performed in general and does not affect the subsequent derivations.
Therefore, we remove the indicator $\tilde{\placeholder}$ in all subsequent derivations.
This rescaling is chosen such that $\tilde{\arr{x}}_k = \diag(\arr{x}_k)\inverse\arr{x}_k$ is the vector of ones, and all scaled parameters in the neighborhood of $\tilde{\arr{x}}_k$ are likewise of the order $1$.

\subsection{Understanding the Gauss-Newton method and gradient descent as dual projections}
\label{sec:descentstep}

\begin{figure}[H]%
	\sameheight{0.85\linewidth}[figures/gaussnewton/]{residual,projection}%
	\centering%
	\hspace*{0.01\linewidth}%
	\subcaptionbox{\label{subfig:gaussnewton:method:residual}}%
	{%
		\includegraphics[height=\samesubheight]{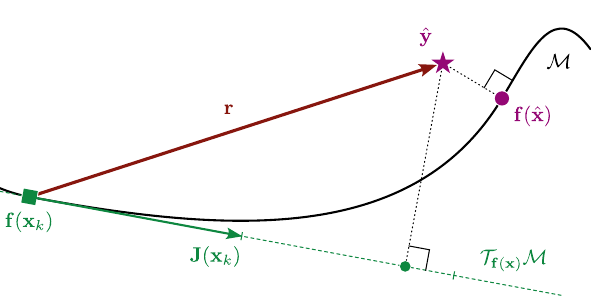}%
	}%
	\hfill%
	\subcaptionbox{\label{subfig:gaussnewton:method:projection}}%
	{%
		\includegraphics[height=\samesubheight]{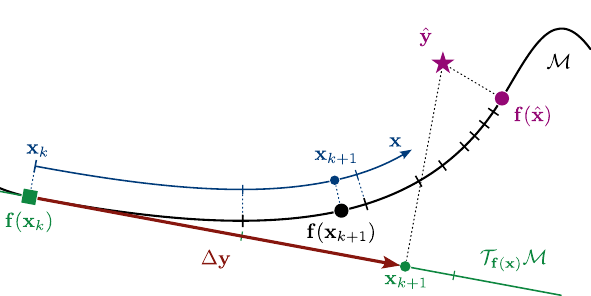}%
	}%
	\hspace*{0.01\linewidth}%
	\captionsetup{width=\linewidth}
	\caption%
	{%
		Geometric interpretation of the Gauss-Newton method via a simplified example.
		The model manifold $\mathcal{M} \in \reals^2$~(\textcolor{bl}{$-$}) is shown for parametrization $\arr{x} \in \reals^1$.
		Model evaluation $\arr{f}(\hat{\arr{x}})$~(\textcolor{pr}{$\bullet$}) at global minimizer $\hat{\arr{x}}$ is the closest projection of reference $\hat{\arr{y}}$~(\textcolor{pr}{$\star$}) onto $\mathcal{M}$.
		At $\arr{f}(\arr{x}_k)$~(\textcolor{gr}{\tiny$\blacksquare$}), $\mathcal{T}_{\arr{f}(\arr{x})}\mathcal{M}$ lies tangent to $\mathcal{M}$, spanned by column vectors of $\arr{J}(\arr{x}_k)$ in \subfigref{gaussnewton:method}{residual}.
		Residual vector $\arr{r}$~(\textcolor{rd}{$\rightarrow$}) can be orthogonally projected onto $\mathcal{T}_{\arr{f}(\arr{x})}\mathcal{M}$ resulting in $\Delta\arr{y}$~(\textcolor{rd}{$\rightarrow$}).
		Projection into $\mathcal{T}_{\arr{f}(\arr{x})}\mathcal{M}$ leads to step $\Delta\arr{x}_k$~(\textcolor{bl}{$\rightarrow$}) along the surface of $\mathcal{M}$ in \subfigref{gaussnewton:method}{projection}.
	}%
	\label{fig:gaussnewton:method}
\end{figure}

In this section, we utilize the geometrical interpretation of the model manifold and Jacobian matrix to derive the Gauss-Newton method defined by \Eqref{optimization:iteration} with step
\begin{equation}\label{eq:optimization:projection:gauss}
	\Delta\arr{x}_k
	=
	\big(\arr{J}(\arr{x}_k)\transpose\arr{J}(\arr{x}_k)\big)\inverse \arr{J}(\arr{x}_k)\transpose\big( \arr{y} - \arr{f}(\arr{x}_k) \big)
	\punct.
\end{equation}
We begin with the simple observation that the column vectors of $\arr{J}(\arr{x})$ in \Eqref{definition:jacobian} form a basis with basis vectors $\arr{e}_i$
\begin{equation}
	\arr{e}_i = \pdv{}{x_i}\arr{f}(\arr{x})
	\punct,
\end{equation}
which spans a Euclidean $m$-dimensional subspace of the ambient signal space centered at $\arr{f}(\arr{x})$.
Generally, the basis vectors $\arr{e}_i$ are neither orthonormal nor orthogonal.
This subspace is called the local tangent space $\mathcal{T}_{\arr{f}(\arr{x})}\mathcal{M}$ and is, as the name implies, tangent to $\mathcal{M}$ at $\arr{f}(\arr{x})$, as shown in \Subfigref{gaussnewton:method}{residual}.
Since the basis vectors $\arr{e}_i$ stem from the partial derivatives of the forward model, any deviation $\Delta\arr{x}$ from $\arr{x}$ can be interpreted as local coordinates with origin at $\arr{x}$ in $\mathcal{T}_{\arr{f}(\arr{x})}\mathcal{M}$.
Subsequently, $\Delta\arr{x}$ can be projected on, or embedded into the ambient signal space via the linear transformation
\begin{equation}
	\Delta\arr{y} = \arr{J}(\arr{x}) \Delta\arr{x}
	\punct,\quad\text{with}\quad
	\arr{f}(\arr{x}) + \Delta\arr{y} \in \mathcal{T}_{\arr{f}(\arr{x})}\mathcal{M}
	\punct.
\end{equation}
In other words, there exists a one-to-one correspondence between the parameter deviations $\Delta\arr{x}$ and signal deviations $\Delta\arr{y}$.
The basis $\arr{e}_i$ can be used to orthogonally project the residual~${\arr{r} = \hat{\arr{y}} - \arr{f}(\arr{x})}$ onto $\mathcal{T}_{\arr{f}(\arr{x})}\mathcal{M}$ to determine the corresponding $\Delta\arr{y}$ s.t.\ ${\forall \arr{e}_i : \arr{r} - \Delta\arr{y} \perp \arr{e}_i}$.
Likewise, the residual can be projected \emph{into} the local coordinate system spanned by $\arr{e}_i$ via the pseudo-inverse of $\arr{J}$ as shown in \Subfigref{gaussnewton:method}{projection}
\begin{equation}
	\Delta\arr{x} = \big(\arr{J}(\arr{x})\transpose\arr{J}(\arr{x})\big)\inverse \arr{J}(\arr{x})\transpose \arr{r}
	\punct.
\end{equation}
This is a generalization of the more commonly known orthogonal vector projection, and directly leads to \Eqref{optimization:projection:gauss}.
This concludes the geometric derivation of the Gauss-Newton method.
It is similar in style to the derivation of the linear least-squares method~\cite{Kern2016}, however, it requires the definition of $\mathcal{T}_{\arr{f}(\arr{x})}\mathcal{M}$ for the linear projection.

Because the pseudo-inverse of the Jacobian acts as a linear map w.r.t.\ the parameters $\arr{x}$, the projected residual is converted into a step inheriting a meaningful distance in the parameter space.
In differential geometry terms, the metric tensor $\arr{G} = \arr{J}\transpose\arr{J}$ correctly converts distances from the ambient signal space into the parameter space~\cite{Tarantola2005}.
A more hands-on but simpler description is that the metric tensor correctly converts the units between spaces.
Furthermore, $\mathcal{T}_{\arr{f}(\arr{x})}\mathcal{M}$ is locally homeomorphic to $\mathcal{M}$, which means that as the distance to the minimizer $\hat{\arr{x}}$ decreases, the projection $\Delta\arr{x}$ becomes a better step estimate.
As such, close to a local minimizer, no vanishing gradient phenomenon occurs as it does for steepest descent methods.
Although in some settings, the Gauss-Newton method may approach saddle points~\cite{Strutz2010}, it does not for least squares~\cite{Bjorck2024}.
This ensures that it converges superlinearly.

\begin{figure}[H]%
	\sameheight{0.85\linewidth}[figures/gaussnewton/]{low-curvature,high-curvature}%
	\centering%
	\hspace*{0.01\linewidth}%
	\subcaptionbox{\label{subfig:gaussnewton:curvature:low}}%
	{%
		\includegraphics[height=\samesubheight]{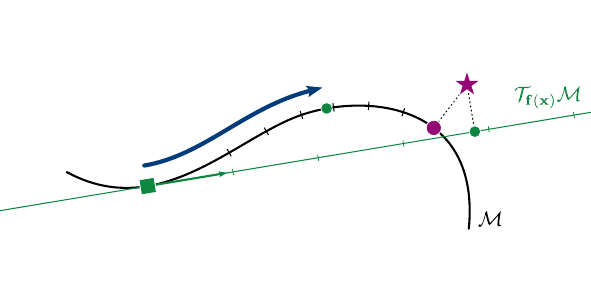}%
	}%
	\hfill%
	\subcaptionbox{\label{subfig:gaussnewton:curvature:high}}%
	{%
		\includegraphics[height=\samesubheight]{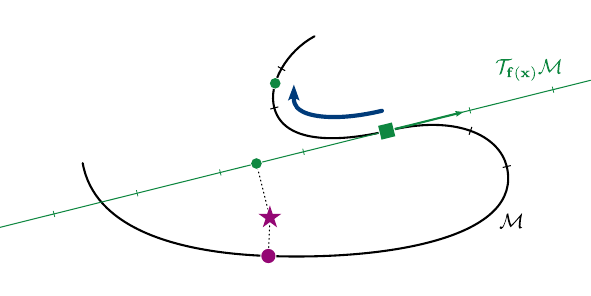}%
	}%
	\hspace*{0.01\linewidth}%
	\captionsetup{width=\linewidth}
	\caption%
	{%
		Weak curvature between a current estimate~(\textcolor{gr}{\tiny$\blacksquare$}) and the true minimizer~(\textcolor{pr}{$\bullet$}) does not affect the direction~(\textcolor{bl}{$\rightarrow$}) of the next estimate~(\textcolor{gr}{$\bullet$}) in \subfigref{gaussnewton:curvature}{low}.
		Strong curvature may cause the residual to project onto the tangent space with the opposite orientation~(\textcolor{bl}{$\leftarrow$}), causing the next estimate to be located arbitrarily far from the true minimizer in \subfigref{gaussnewton:curvature}{high}.
		The true minimizer is located at the closest projection of the reference~(\textcolor{pr}{$\star$}) onto the manifold.
		The next estimate is found by the orthogonal projection of the residual between the current estimate and reference onto the tangent space.
	}%
	\label{fig:gaussnewton:curvature}
\end{figure}

The Gauss-Newton method always approaches the global minimizer given that the initial estimate is sufficiently close to it, such that the extrinsic curvature does not bend the model manifold to such a degree that the orientation of the projection is reversed.
This can be understood geometrically via a comparison of \Subfigsref{gaussnewton:curvature}{low}~and~\subfigref{gaussnewton:curvature}{high}, where the residual vector may be oriented opposite to the path that should be taken in the parameter space owing to how the model manifold with its extrinsic curvature is embedded in the signal space in relation to the orientation of its coordinate lines.
Then, the local (Euclidean) approximation via the tangent space is insufficient to capture the global behavior of the extrinsic embedding, i.e., straight paths in the parameter space do not concur with the shortest paths (geodesics) along the manifold surface.
All local derivative-based optimization methods suffer from this problem, which cannot be alleviated without additional global information. (Typically, line search is recommended.)

\begin{figure}[H]%
	\sameheight{0.7\linewidth}[figures/projectionssimple/]{tangent,projections,levelsets}%
	\centering%
	\hspace*{\stretch{0.5}}%
	\subcaptionbox{\label{subfig:manifold:space:tangent}}%
	{%
		\includegraphics[height=\samesubheight]{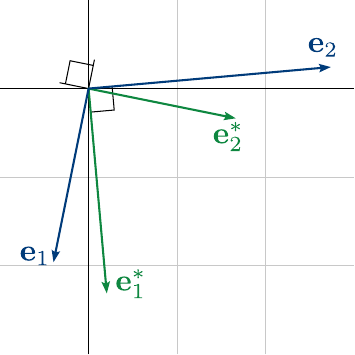}%
	}%
	\hspace*{\stretch{1}}%
	\subcaptionbox{\label{subfig:manifold:space:projections}}%
	{%
		\includegraphics[height=\samesubheight]{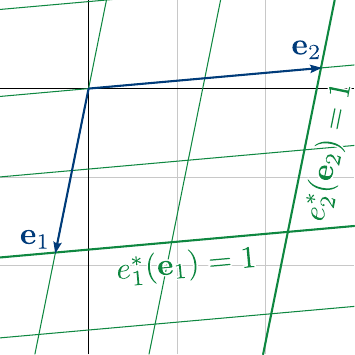}%
	}%
	\captionsetup{width=0.8\linewidth}
	\hspace*{\stretch{1}}%
	\subcaptionbox{\label{subfig:manifold:space:levelsets}}%
	{%
		\includegraphics[height=\samesubheight]{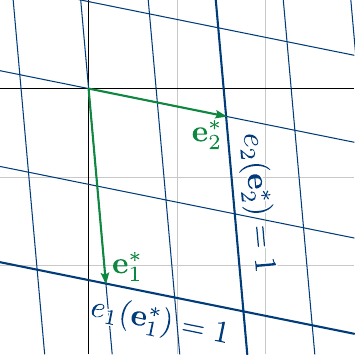}%
	}%
	\hspace*{\stretch{0.5}}%
	\captionsetup{width=\linewidth}
	\caption%
	{%
		Basis vectors $\arr{e}_i$ and dual vectors $\arr{e}_i^\ast$ which span $\mathcal{T}_{\arr{f}(\arr{x})}\mathcal{M}$ \subfigref{manifold:space}{tangent}, functionals $e_i^\ast(\updelta\arr{y})$ \subfigref{manifold:space}{projections}, and dual functionals $e_i(\updelta\arr{y})$ \subfigref{manifold:space}{levelsets}.
	}%
	\label{fig:manifold:space}
\end{figure}

We now focus on the dual space representation of the Jacobian~$\arr{J}$ to obtain the standard gradient descent method and subsequently derive our geometrically motivated automatic step adaption.
Once one recognizes that the basis vectors $\arr{e}_i$ span a subspace of $\reals^n$, similarly a set of functionals ${\updelta x_i = e^\ast_i(\updelta\arr{y})}$ dual to $\arr{e}_i$ can be found that have to fulfill the Kronecker delta property~\cite{Luenberger1997}
\begin{equation}\label{eq:dual:function:kronecker}
	e^\ast_j(\arr{e}_i) = \delta_{ij}
	\punct.
\end{equation}
This set of functionals, due to their nature as linear transformations in finite-dimensional spaces, can be defined via a set of vectors $\arr{e}^\ast_i$ s.t.\
\begin{equation}\label{eq:dual:vectorfunction}
	\updelta x_i = \arr{e}^\ast_i \cdot \updelta\arr{y}
	\punct,
\end{equation}
which equally has to fullfull
\begin{equation}\label{eq:dual:vector:kronecker}
	\arr{e}^\ast_j \cdot \arr{e}_i = \delta_{ij}
	\punct,
\end{equation}
or more compactly using ${\arr{J}, \arr{J}^\ast \in \reals^{n \times m}}$, the Jacobian and the Jacobian dual respectively,
\begin{equation}\label{eq:dual:matrix:kronecker}
	\arr{J}\transpose\arr{J}^\ast = \arr{I}_m
	\punct.
\end{equation}
Usually, the basis vectors~$\arr{e}^\ast_i$ are considered as row-vectors for defining the functionals~$e^\ast_i$.
However, we abuse this notation to utilize $\arr{e}^\ast_i$ as proper vectors, from which the standard formulation of the gradient descent step naturally arises.
Since $\arr{J}$ and $\arr{J}^\ast$ are of the same size, it is not difficult to imagine that the residual vector may be projected onto this "vectorized" dual basis, as in \Eqref{optimization:projection:gauss}, such that
\begin{equation}\label{eq:optimization:projection:dual}
	\Delta\arr{x}^\ast_k = \big(\arr{J}^\ast(\arr{x}_k)\transpose\arr{J}^\ast(\arr{x}_k)\big)\inverse \arr{J}^\ast(\arr{x}_k)\transpose \big(\arr{y} - \arr{f}(\arr{x}_k)\big)
	\punct.
\end{equation}
However, the pseudo-inverse of $\arr{J}^\ast$ does not need be explicitly evaluated since \Eqref{dual:matrix:kronecker} directly provides the inverse as follows
\begin{equation}
	(\arr{J}^\ast)\inverse = \arr{J}\transpose
	\punct,
\end{equation}
such that \Eqref{optimization:projection:dual} simplifies to
\begin{equation}\label{eq:optimization:projection:gradientdescent}
	\Delta\arr{x}^\ast_k = \arr{J}(\arr{x}_k)\transpose\big( \arr{y} - \arr{f}(\arr{x}_k) \big)
	\punct.
\end{equation}
\Eqref*{optimization:projection:gradientdescent} is precisely the gradient descent step, compare with \Eqref{gradient:loss}.
\Figref*{manifold:space} shows a comparison of the separate bases and projective planes with the different coordinate lines they produce.

From this, it follows directly that the gradient descent algorithm cannot possibly convert distances correctly, and the step $\Delta\arr{x}^\ast_k$ found does not correspond to a step in the parameter space but its dual.
The gradient descent step only points in the direction of the steepest descent for the level set functions~${e^\ast_i(\updelta\arr{y}) - \updelta x_i = 0}$~\cite{Hansen2013}.
These define coordinate lines orthogonal to the vectors $\arr{e}^\ast_i$ and are equal to one where they meet their respective $\arr{e}_i$.
This leads to the equivalence of \Eqsref{optimization:projection:gauss}~and~\eqref{dual:vectorfunction} assuming ${\updelta\arr{y} = \arr{y} - \arr{f}(\arr{x})}$ and explains why distances are meaningful w.r.t.\ $\arr{e}_i$ and hence the parameter space.
The same is not true for gradient descent, which would be equivalent to the "dual dual" function basis using $\arr{e}_i$
\begin{equation}\label{eq:dualdual:vectorfunction}
	\updelta x^\ast_i = \arr{e}_i \cdot \updelta\arr{y}
	\punct,
\end{equation}
and defines distances w.r.t.\ $\arr{e}^\ast_i$, i.e., \Eqsref{optimization:projection:gradientdescent}~and~\eqref{dualdual:vectorfunction} are likewise equivalent.
The gradient descent step can then be understood as being reciprocal to the Gauss-Newton step and, as such, \emph{always} requires correction of its length~\cite{Tarantola2005}.

Therefore, we argue that it is not only practically justified to perform a step size adaption of the gradient descent method but geometrically essential to rescale the descent vector.
In the Gauss-Newton method, this is handled implicitly via the metric tensor.
In the context of the Levenberg-Marquardt method, instead of $\arr{I}_m$ in \Eqref{optimization:projection:levenberg}, the use of $\diag(\arr{G})$ to implicitly rescale the gradient descent step is common~\cite{Nocedal2006}.
This makes sense from the perspective that each individual component is rescaled according to its own order of magnitude.
However, we highlight that both $\arr{G}$ and $\diag(\arr{G})$ fundamentally change the steepest descent direction.
We argue that it is precisely the pure downhill direction that the gradient provides, which leads to more stable convergence far away from the minimizer.
Therefore, we intend to apply only a purely scalar rescaling of the step to preserve the original direction.
This is further justified because we have already equalized the discrepancies in the orders of magnitude via \Eqref{jacobian:rescale}.
We propose the following nonlinear formulation of a step adaption~$\lambda_k$, motivated by bounding the gradient descent step by the Gauss-Newton step via the metric distance function $g_{\arr{x}}(\Delta\arr{x})$ using the metric tensor
\begin{equation}\label{eq:scaling:nonlinear}
	\lambda_k
	=
	\frac{g_{\arr{x}}(\Delta\arr{x}_k)}{g_{\arr{x}}(\Delta\arr{x}^\ast_k)}
	\punct,\quad
	g_{\arr{x}}(\Delta\arr{x}) = \sqrt{\Delta\arr{x}\transpose\arr{G}(\arr{x})\Delta\arr{x}}
	\punct,
\end{equation}
under the assumption that the Gauss-Newton step may not point in the optimal direction, but scales the step optimally~\cite{Bjorck2024}.
\Eqref*{scaling:nonlinear} can be further simplified by substituting ${\Delta\arr{x}_k = \arr{G}(\arr{x}_k)\inverse \Delta\arr{x}^\ast_k}$ in \Eqref{scaling:nonlinear}, compare \Eqsref{optimization:projection:gauss}~and~\eqref{optimization:projection:gradientdescent},
\begin{equation}\label{eq:scaling:nonlinear:simplified}
	\lambda_k
	=
	\sqrt{\frac{(\Delta\arr{x}^\ast_k)\transpose\arr{G}\inverse\Delta\arr{x}^\ast_k}{(\Delta\arr{x}^\ast_k)\transpose\arr{G}\Delta\arr{x}^\ast_k}}
	\punct,
\end{equation}
which can then be directly computed via the gradient descent step $\Delta\arr{x}^\ast_k$ and yields the novel locally step-size-adapted gradient descent for least-squares problems
\begin{equation}\label{eq:optimization:projection:gradientdescent:corrected}
	\arr{x}_{k+1}
	=
	\arr{x}_k + \lambda_k \Delta\arr{x}^\ast_k
	\punct.
\end{equation}
This scaling factor nonlinearly transforms the gradient descent step to be bounded by the associated Gauss-Newton step.

Gradient descent cannot show the same convergence rates as the Gauss-Newton method, and must be considered a purely steepest descent method for a given objective function.
This behavior of the gradient descent method is all too familiar to practitioners, which is why it is widely considered to provide only a search direction without offering a meaningful step length, additionally necessitating a line search to prove convergence.
This explanation offers a more fundamental reason for why this behavior arises in nonlinear least squares. It also highlights a specific approach to address this issue, without the explicit need for a line search or additional hyperparameters.

\subsection{Automatic step size for Levenberg-Marquardt method}

Regardless of the previously described limitations of gradient descent, it has one major advantage over the Gauss-Newton method.
It tends to exhibit greater stability even when (relatively) far from a minimizer compared with the Gauss-Newton method.
One explanation for this may be that it is a pure descent method~\cite{Strutz2010}.
This led to the development of the Levenberg-Marquardt method, which seeks to unify the best aspects of both methods: stability far from the minimizer and rapid convergence when approaching it.
Note the major difference when comparing \Eqsref{optimization:projection:gauss}~and~\eqref{optimization:projection:gradientdescent}, namely, the left-multiplication by the inverse metric tensor.
The simple addition and introduction of a scaling factor $\eta$ for the gradient descent step then yields the Levenberg-Marquardt step~\cite{Levenberg1944,Marquardt1963,Nocedal2006}
\begin{equation}\label{eq:optimization:projection:levenberg}
	\Delta\arr{x}_k
	=
	\big( \arr{J}(\arr{x}_k)\transpose\arr{J}(\arr{x}_k) + \eta \, \arr{I}_m \big)\inverse \arr{J}(\arr{x}_k)\transpose\big( \arr{y} - \arr{f}(\arr{x}_k) \big)
	\punct.
\end{equation}

We propose the following automatic scaling scheme, considering the initial residual ${\Delta\arr{y}_0 = \arr{y} - \arr{f}(\arr{x}_0)}$ and computing the ratio of all subsequent residuals ${\Delta\arr{y}_k = \arr{y} - \arr{f}(\arr{x}_k)}$ as
\begin{equation}\label{eq:automatic:adaption}
	\bar{\eta}_k
	=
	\frac{\norm{\Delta\arr{y}_k}}{\norm{\Delta\arr{y}_0}}
	\punct,\quad
	\lim_{\arr{x}_k \rightarrow \hat{\arr{x}}} \bar{\eta}_k \le 1
	\punct,
\end{equation}
assuming that $\Delta\arr{x}_k$ strictly points in a descent direction towards a minimizer.
Combining \Eqsref{scaling:nonlinear:simplified}, \eqref{optimization:projection:gradientdescent:corrected}, and \eqref{optimization:projection:levenberg}
\begin{equation}
	\Delta\arr{x}_k
	=
	\big( \arr{J}(\arr{x}_k)\transpose\arr{J}(\arr{x}_k) + \bar{\eta}_k \,  \lambda_k\inverse \arr{I}_m \big)\inverse \Delta\arr{x}^\ast_k
	\punct,\quad
	\Delta\arr{x}^\ast_k
	=
	\arr{J}(\arr{x}_k)\transpose(\arr{y} - \arr{f})
\end{equation}
gives rise to our proposed automatic step-size-adapted Levenberg–Marquardt method with $\eta = \bar{\eta}_k \lambda_k\inverse$.

\begin{figure}[H]%
	\sameheight{0.9\linewidth}[figures/optimizationmethod/]{AmbientSpace,GradientDescent,GaussNewton,Ours}%
	\centering%
	\subcaptionbox{\label{subfig:optimizationmethods:ambientspace}}%
	{%
		\includegraphics[height=\samesubheight]{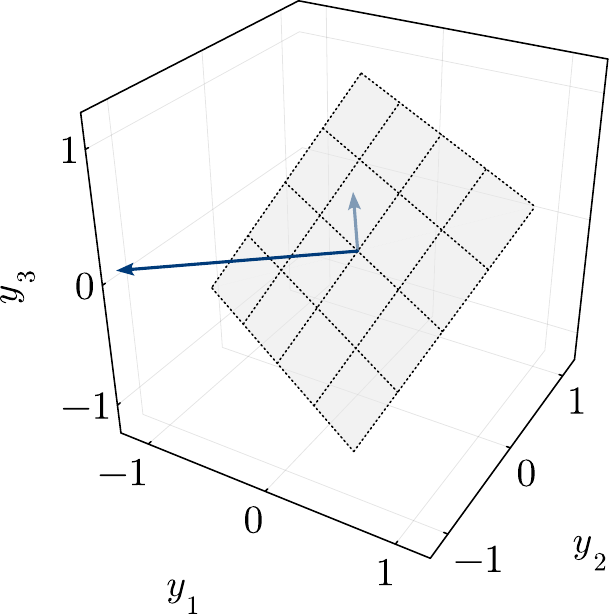}%
		\vspace*{-5mm}
	}%
	\hfill%
	\subcaptionbox{\label{subfig:optimizationmethods:gradientdescent}}%
	{%
		\includegraphics[height=\samesubheight]{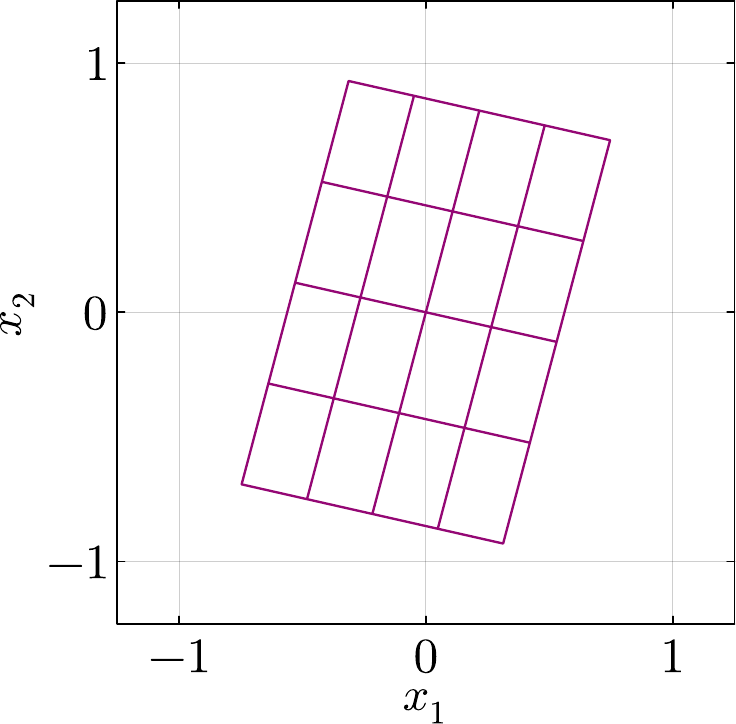}%
		\vspace*{-5mm}
	}%
	\hfill%
	\subcaptionbox{\label{subfig:optimizationmethods:gaussnewton}}%
	{%
		\includegraphics[height=\samesubheight]{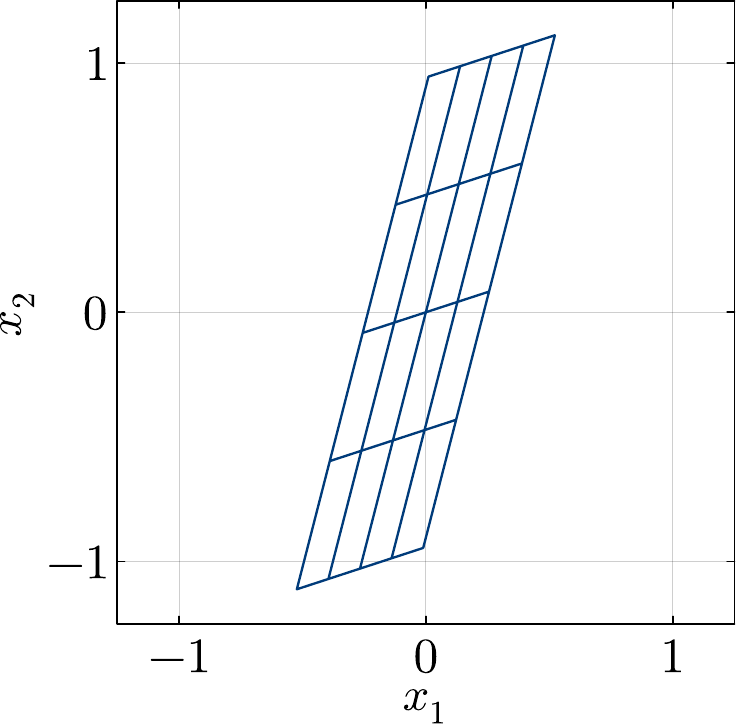}%
		\vspace*{-5mm}
	}%
	\hfill%
	\subcaptionbox{\label{subfig:optimizationmethods:ours}}%
	{%
		\includegraphics[height=\samesubheight]{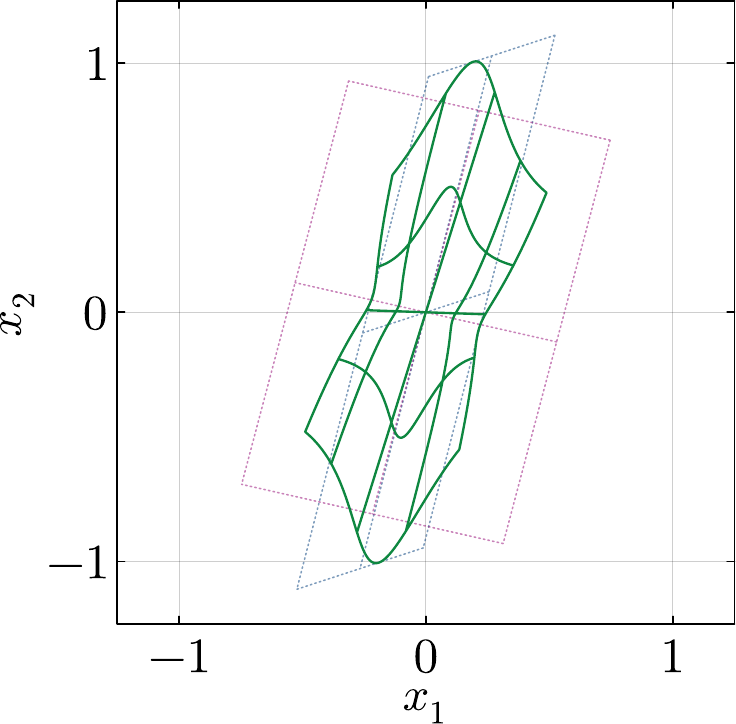}%
		\vspace*{-5mm}
	}%
	\captionsetup{width=\linewidth}
	\caption%
	{%
		Uniformly distributed residuals belonging to grey surface projected onto the tangent space spanned by basis vectors~(\textcolor{bl}{$\rightarrow$}) in~\subfigref{optimizationmethods}{ambientspace}: for gradient descent~\subfigref{optimizationmethods}{gradientdescent}, Gauss-Newton~\subfigref{optimizationmethods}{gaussnewton}, and modified Levenberg-Marquardt with $\bar{\eta} = 1$~\subfigref{optimizationmethods}{ours}.
		Note that the projected residuals in our method are bounded by the projections of the Gauss-Newton method.
	}%
	\label{fig:optimizationmethods}
\end{figure}

The relative interpolation factor $\bar{\eta}_k$ in \Eqref{automatic:adaption} is primarily motivated by earlier considerations involving \Eqsref{jacobian:rescale}~and~\eqref{scaling:nonlinear:simplified}.
Provided a strictly decreasing sequence of $\arr{x}_k$, which is not guaranteed, $\bar{\eta}_k \in [0, 1]$ is bounded.
Obviously, $\bar{\eta}_k$ may grow larger than one for unintended uphill directions\footnote{In nonlinear problems, there is no guarantee that $\arr{f}(\arr{x}_{k+1}) > \arr{f}(\arr{x}_{k})$ unless a line search algorithm is used to enforce this condition.}.
This is not problematic in and of itself, as other methodologies allow larger step adaptions as well~\cite{Yamashita2001,Fan2005,Ma2007,Fan2009,Fan2012,Fan2019,Li2023b,Karas2016,Amini2018,Li2023a,Han2024}.
However, we do not require any drastic modifications to the step size because our method is designed to provide a correct rescaling via $\lambda$.
The factor $\bar{\eta}_k$ merely serves the purpose of continuously transitioning to the Gauss-Newton method to achieve rapid convergence to the local minimizer.
In a non-convex setting, $\bar{\eta}_k > 0$ if an incorrect local minimizer is approached.
See~\Figref{optimizationmethods} for a comparison of different steps in a parameter space $\arr{x} \in \reals^2$ for a plane of residuals $\arr{r} \in \reals^3$ given a random Jacobian.

\subsection{Novel objective function on the basis of an autocorrelation of the frequency content}
\label{sec:objectivefunction:novel}

The envelope of signals has previously been shown to be an effective transformation for improving the objective function surfaces of inverse problems~\cite{Kristekova2006,Kristekova2009,Itner2021PAMMb,Bulling2022}.
This is likewise demonstrated in \Subfigsref{model:manifold}{signals}~and~\subfigref{model:manifold}{envelopes}; the envelope unwraps the model manifold that is otherwise characterized by a tight spiral loop about the origin.
Nonetheless, the envelope is not sufficient to transform the model output to facilitate optimization over a broad parameter range; therefore, it cannot be expected to work well for more complicated material symmetries, including damping.
The objective function that arises from a modification of the Fourier coefficients, as presented here, leads to a cost surface that further assists in optimization towards the global minimizer.
We first present the derivation and novel objective function, and then compare it with alternatives to highlight its improvements via example objective surfaces for isotropic material symmetry.

As mentioned in \Secref{problem}, the forward model is solved only for a certain subset of Fourier coefficients of spatial and temporal modes to reduce the compute time of a simulation.
Therefore, the corresponding response signal in time contains substantially less information than that suggested by the number of time steps (on the order $2^{12}$).
Thus, it naturally follows that it may be reasonable to compare signals directly in the complex frequency space.
This argument provides the initial motivation for the following derivation.

The envelope $e(t)$ of signal $u(t)$ is based on the analytic signal $a(t)$, where $\mathcal{H}(u)(t)$ is the Hilbert transform of $u(t)$ evaluated at $t$~\cite{Cohen1995}
\begin{equation}\label{eq:envelope}
	e(t) = \abs*{a(t)} = \sqrt{ u^2(t) + \mathcal{H}^2(u)(t) }
	\punct,\quad
	a(t) = u(t) + \imag \mathcal{H}(u)(t)
	\punct.
\end{equation}
The square root in \Eqref{envelope} entails that the Fourier transform of the envelope is not trivially related to that of the signal.
Instead, the square of the envelope
\begin{equation}\label{eq:envelope:square}
	e^2(t) = u^2(t) + \mathcal{H}^2(u)(t)
\end{equation}
can be considered because $\forall t \ge 0 : e(t) \ge 0$ such that squaring the envelope of the signal does not change any information and, in fact, can be simplified in the frequency domain.
This is due to the inverse relationship of multiplication and the convolution operation~($\conv$) between the time and frequency domains, leading to
\begin{equation}\label{eq:envelope:square:convolution}
	E(\omega) = U(\omega) \conv U(\omega) + \mathcal{H}(U)(\omega) \conv \mathcal{H}(U)(\omega)
	\punct.
\end{equation}
Here, $E$, $U$, and $H$ refer to the Fourier coefficients w.r.t.\ the angular frequency $\omega \in \reals$
\begin{align}
	E(\omega)
	&=
	\int_{-\infty}^{+\infty} e^2(t) \, \euler^{-2\pi \imag \omega t} \dd{t}
	\punct,
	&
	U(\omega)
	&=
	\int_{-\infty}^{+\infty} u(t) \, \euler^{-2\pi \imag \omega t} \dd{t}
	\punct,
	&
	\mathcal{H}(U)(\omega)
	&=
	\int_{-\infty}^{+\infty} \mathcal{H}(u)(t) \, \euler^{-2\pi \imag \omega t} \dd{t}
	\punct.
\end{align}
Because we strictly apply signals containing a finite amount of energy, we can expect any coefficient $U(\omega = 0) = 0$.
As such, we restrict $\omega \in \reals \setminus \{0\}$ to simplify the following derivation.

We now address each summand of \Eqref{envelope:square:convolution} individually, beginning with the second summand.
The coefficients of the Hilbert transform are directly related to the coefficients of the signal itself through
\begin{align}\label{eq:hilbert:sgn}
	\mathcal{H}(U)(\omega) &= -\imag \sgn(\omega) U(\omega)
	\punct,
	&
	\sgn(\omega) &= \begin{cases} -1 \text{ , } \omega < 0 \\ \sm 1 \text{ , } \omega \ge 0 \end{cases}
	\punct.
\end{align}
We express the $\sgn$-function as a decomposition via the Heaviside step function $H$
\begin{align}\label{eq:step}
	\sgn(\omega) &= H(\omega) - H(-\omega)
	\punct,
	&
	H(\omega) &= \begin{cases} 0 &\text{, } \omega < 0 \\ 1 &\text{, } \omega \ge 0 \end{cases}
	\punct.
\end{align}
This allows us to further decompose the Fourier transform $U$ into components only associated with the positive and negative frequencies, respectively, since $\forall \omega < 0: H(\omega) = 0$ and $\forall \omega > 0: H(-\omega) = 0$
\begin{equation}\label{eq:step:explicit}
	\sgn(\omega) U(\omega)
	=
	H(\omega) U(\omega) - H(-\omega) \bar{U}(\omega) 
	\punct,
\end{equation}
where $\bar{U}(\omega)$ are the complex conjugate coefficients of the negative frequency range, with $U(-\omega) = \bar{U}(\omega)$.
We introduce ${U^+ = H(\omega) U(\omega)}$ and ${\bar{U}^- = H(-\omega) \bar{U}(\omega)}$ for conciseness.
Substitution of \Eqref{step:explicit} in \Eqref{hilbert:sgn} then leads to
\begin{equation}\label{eq:hilbert:parts}
	\mathcal{H}(U)(\omega)
	=
	\imag (U^- - U^+)
	\punct,
\end{equation}
Finally, the autoconvolution in \Eqref{envelope:square:convolution} can be separated into positive and negative frequency components using \Eqref{hilbert:parts}.
We use the algebraic properties of the convolution operator, namely, associativity with scalar multiplication, commutativity, and distributivity
\begin{align*}\label{eq:hilbert:convolution}
	\mathcal{H}(U)(\omega) \conv \mathcal{H}(U)(\omega)
	&=
	\imag \left( \bar{U}^- - U^+ \right) \conv \imag \left( \bar{U}^- - U^+ \right)
	\\
	&=
	2 (U^+ \conv \bar{U}^-) - (U^+ \conv U^+) - (\bar{U}^- \conv \bar{U}^-) \donumber
	\punct.
\end{align*}
The first summand in \Eqref{envelope:square:convolution} can be manipulated in a similar fashion to cancel terms by using the identity property ${\forall \omega \ne 0: H(\omega) + H(-\omega) = 1}$
\begin{equation}\label{eq:signal:parts}
	U(\omega)
	=
	\big( H(\omega) + H(-\omega) \big) \, U(\omega)
	=
	U^+ + \bar{U}^-
	\punct,
\end{equation}
and by substituting \Eqref{signal:parts} into the autoconvolution in \Eqref{envelope:square:convolution}, the first summand can be separated similarly
\begin{align*}\label{eq:signal:convolution}
	U(\omega) \conv U(\omega)
	&=
	( U^+ + \bar{U}^- ) \conv ( U^+ + \bar{U}^- )
	\\
	&=
	2 (U^+ \conv \bar{U}^-) + (U^+ \conv U^+) + (\bar{U}^- \conv \bar{U}^-) \donumber
	\punct.
\end{align*}
Adding \Eqref{hilbert:convolution} and \Eqref{signal:convolution} directly leads to
\begin{equation}\label{eq:convolution:final}
	E(\omega) = 4 (U^+ \conv \bar{U}^-)
	\punct,\quad
	\omega \ne 0
	\punct.
\end{equation}

Since the positive and negative frequencies are removed from the respective convolution terms, \Eqref{convolution:final} can be simply restated as the autocorrelation of only the positive frequency content with the cross-correlation operator $(\xcorr)$:
\begin{equation}\label{eq:autocorrelation}
	E(\omega) = 4 \big( U(\omega) \xcorr U(\omega) \big)
	\punct,\quad
	\omega > 0
	\punct.
\end{equation}
Here, the first (second) half of the autocorrelation in \Eqref{autocorrelation} corresponds to the negative (positive) frequency content of the squared envelope.
Furthermore, the information content of these negative and positive frequencies is equivalent because of complex conjugation.
Therefore, only the second half of the autocorrelated positive frequency content is required to establish a comparative measure of the signals.
Additionally, ignoring the constant factor of $4$ (because it is likewise not relevant for a comparative measure), the discrete coefficients $E_k$ for discrete frequencies $\omega_k \ge 0$ of the squared envelope can be computed by iterating over the original set of discrete coefficients $U_i$:
\begin{equation}\label{eq:autocorrelation:coefficients}
	E_k = \sum_{i=k}^{n^+ - 1} U_{i+1} \bar{U}_{i-k+1}
	\punct,\quad
	i,k \in \{0, \dots, n^+ - 1\}
	\punct,
\end{equation}
where $n^+$ is the number of discrete Fourier coefficients associated with the positive frequencies.
As a note, specifically for $k=0$ (which corresponds to the static mode of the envelope \footnote{Irrespective of our assumption that we exclude the static mode of the signal itself because it is zero, the envelope possesses a computable static mode.}), \Eqref{autocorrelation:coefficients} evaluates to $E_0 \in \reals_{>0}$, and corresponds to the mean of the envelope, which must be strictly positive.
Finally, let $\hat{\arr{a}}$ and $\arr{a}$ be the vectors of modified Fourier coefficients of the measurement and simulation, respectively, with their elements defined by \Eqref{autocorrelation:coefficients}.
Then, the following residual can be defined for the unwrapped arguments (phase angles) of the signals in the frequency domain with
\begin{equation}\label{eq:difference:phase:unsafe}
	\arr{r}
	=
	\arg(\hat{\arr{a}}) - \arg(\arr{a})
	\punct,\quad
	\arg(z)
	=
	\atan2(\imagpart(z), \realpart(z))
	\punct,\quad
	z \in \complexs
	\,\text{,}
\end{equation}
where the $\arg$-function is applied element-wise.
We assume that the output of the $\arg$-function for vector inputs is implicitly unwrapped by removing discontinuities between jumps in phases from $+\pi$ to $-\pi$~\cite{Oppenheim2013}.

Applying \Eqref{difference:phase:unsafe} leads to discontinuities in the objective function owing to numerical imprecisions in the computation.
Alternatively, the residuals can be computed as
\begin{equation}\label{eq:difference:complex}
	r_k
	=
	\arg(\hat{a}_k / a_k)
	\punct,
\end{equation}
but this method has similar numerical drawbacks and deviates from the standard formulation of the least-squares objective.
The residual formulation in \Eqref{difference:phase:unsafe} can be retained by utilizing a third fictitious signal\footnote{This methodology was inspired by a discussion on \url{https://math.stackexchange.com/} and then further modified. We regret to inform that a link to the original discussion could not be recovered at time of publication.} with a linear phase w.r.t.\ $\omega$.
The phases of the original signals are normalized w.r.t.\ this fictitious signal, which is designed to make the subsequent comparison of the phases more robust.
Because the Fourier coefficients in $\hat{\arr{a}}$ and $\arr{a}$ are compared in the frequency domain, we introduce substitute values for pseudo-phase $\varphi_k$ and pseudo-coefficient $Y_k$ of this ficticious signal defined by
\begin{align}\label{eq:difference:pseudo}
	\varphi_k
	&=
	\tfrac{1}{2} T \tfrac{2\pi}{T} k
	=
	\pi k
	\punct,
	&
	Y_k
	&=
	\cos(\varphi_k) + \imag \sin(\varphi_k)
	=
	(-1)^{k}
	\punct.
\end{align}
This signal is specifically constructed such that its phase has a slope of $\tfrac{1}{2}$ for a frequency step of $\Delta\omega = \tfrac{2\pi}{T}$ between the frequencies of the discrete Fourier coefficients.
This corresponds to a factor of $\pi$ for the discrete pseudo-phases $\varphi_k$ as shown in \Eqref{difference:pseudo}.
This makes it possible to indirectly apply \Eqref{difference:complex} by first normalizing each individual coefficient with
\begin{equation}\label{eq:difference:normalize}
	\arg\subrm{normalized}(E_k) = \arg(E_k / Y_k) - \varphi_k
	\punct.
\end{equation}
Empirically, we find this to be the numerically most robust comparison between phases.

In addition, the phase correlations associated with frequencies outside the excitation range are numerically unstable.
The associated amplitudes become miniscule (where rapidly $\lim_{\omega \rightarrow \infty} |E(\omega)| = 0$), leading to undefined phases, and hence, a numerically unstable evaluation of the same.
We apply the numerical damping coefficients $\gamma_k$ derived from the excitation signal in \Eqref{excitation}
\begin{equation}
	\gamma_k
	=
	\exp(-C \, k^2 / (b \, T)^2)
\end{equation}
with $C \in [1, 10]$ to circumvent the numerical instability, and establish the stable argument
\begin{equation}
	\arg\subrm{stable}(E_k)
	=
	\gamma_k \arg\subrm{normalized}(E_k)
	\punct.
\end{equation}
Then, \Eqref{difference:phase:unsafe} can be rephrased to
\begin{equation}\label{eq:difference:phase:safe}
	\arr{r}
	=
	\arg\subrm{stable}(\hat{\arr{a}}) - \arg\subrm{stable}(\arr{a})
	\,\text{.}
\end{equation}

\begin{figure}[H]%
	\sameheight{0.9\linewidth}[figures/costsurface/]{signal,envelope,autocorrelation}%
	\centering%
	\hspace*{0.01\linewidth}%
	\begin{subcaptionblock}{0.3\linewidth}%
		\subcaption{\label{subfig:costsurface:signal}}%
		\includegraphics[height=\samesubheight]{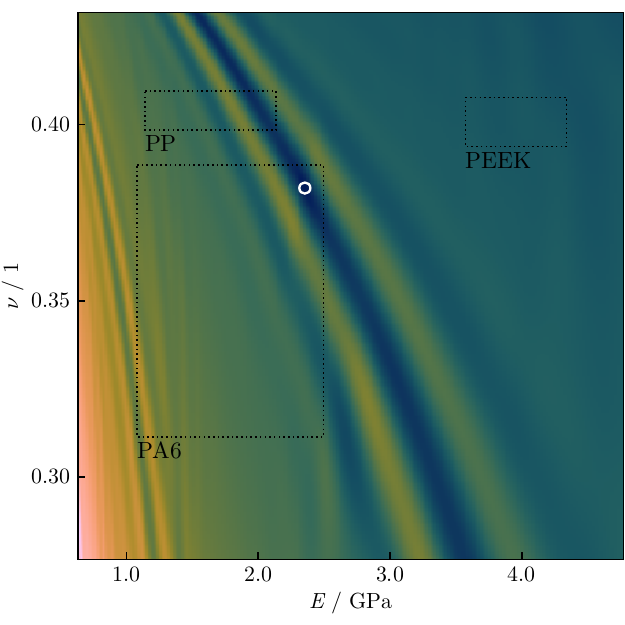}%
	\end{subcaptionblock}%
	\hfill%
	\begin{subcaptionblock}{0.3\linewidth}%
		\subcaption{\label{subfig:costsurface:envelope}}%
		\includegraphics[height=\samesubheight]{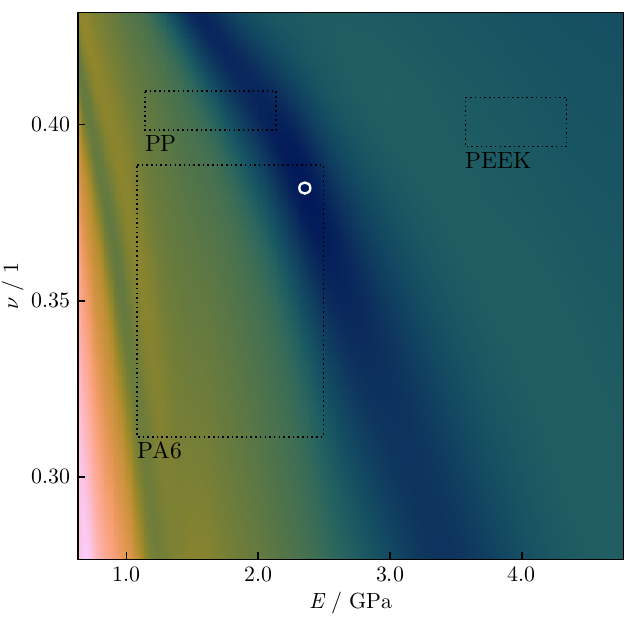}%
	\end{subcaptionblock}%
	\hfill%
	\begin{subcaptionblock}{0.3\linewidth}%
		\subcaption{\label{subfig:costsurface:autocorrelation}}%
		\includegraphics[height=\samesubheight]{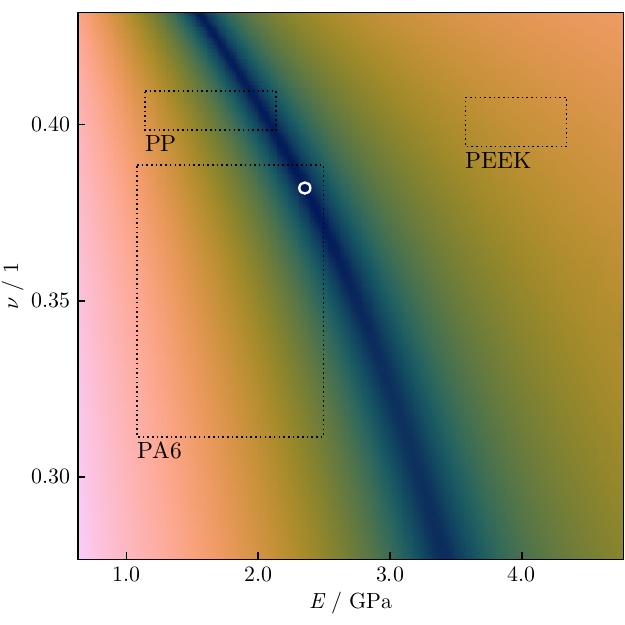}%
	\end{subcaptionblock}%
	\hspace*{0.01\linewidth}%
	\captionsetup{width=\linewidth}
	\caption%
	{%
		Surface of least-squares objective function for signals~\subfigref{costsurface}{signal}, envelopes~\subfigref{costsurface}{envelope}, and autocorrelated~phases~\subfigref{costsurface}{autocorrelation} (with $C = 1$) in the isotropic parameter space containing two standard deviations of the marginal distributions of parameters for PEEK, PA6, and PP.
		Boxes wrap one standard deviation of each marginal distribution.
		The minimizer is indicated by ($\circ$).
	}%
	\label{fig:costsurface}
\end{figure}

A comparison of different objective function surfaces for a particular reference is shown in \Figref{costsurface}.
Whereas the objective function surface for signals is riddled with local optima, the envelope already leads to a great reduction in the number of suboptimal optima within two standard deviations of the marginal distributions\footnote{These are established in \Secref{analysis:materialbounds}.} of the materials of interest.
The residuals, based on autocorrelation, completely convexify the objective function within the search space.

\begin{figure}[H]%
	\sameheight{0.85\linewidth}[figures/objectivefunction/]{signal,envelope}%
	\centering%
	\includegraphics[height=0.15\samesubheight]{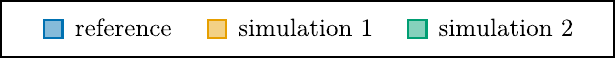}\\%
	\hspace*{0.03\linewidth}%
	\begin{subcaptionblock}{0.425\linewidth}%
		\subcaption{\label{subfig:objectivefunction:signal}}%
		\includegraphics[height=\samesubheight]{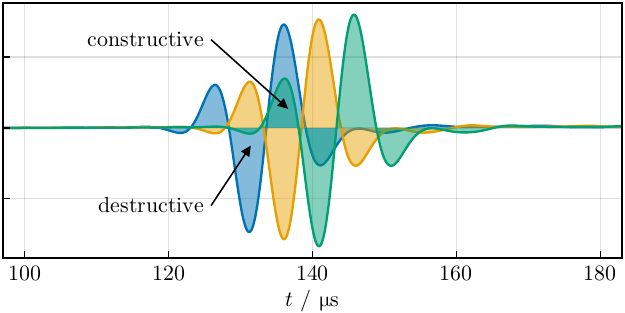}%
	\end{subcaptionblock}%
	\hfill%
	\begin{subcaptionblock}{0.425\linewidth}%
		\subcaption{\label{subfig:objectivefunction:envelope}}%
		\includegraphics[height=\samesubheight]{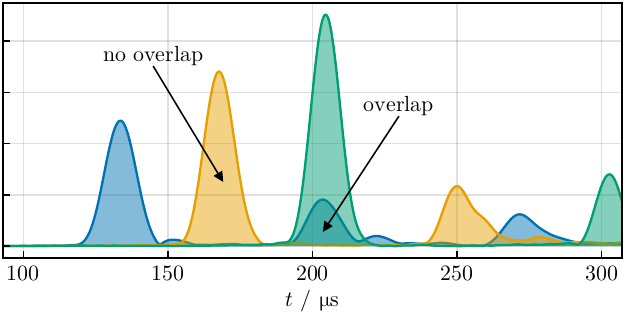}%
	\end{subcaptionblock}%
	\hspace*{0.03\linewidth}%
	\captionsetup{width=\linewidth}
	\caption{%
		Overlap in signals~\subfigref{objectivefunction}{signal} and envelopes~\subfigref{objectivefunction}{envelope} for different combinations of $E$ and $\nu$ located at the local minima and maxima of the respective objective functions.
		For signals, the first wave packet is affected by the constructive and destructive interference of oscillations.
		For envelopes, the arrival times of individual wave packets may overlap.
	}%
	\label{fig:objectivefunction}%
\end{figure}

We now discuss how the transformation and residual proposed in \Eqref{difference:phase:safe} leads to  a greater region of convexity.
Minor shifts in the arrival time of the primary wave packet lead to constructive and destructive interference of the underlying oscillations of the signals, as shown in \Subfigref{objectivefunction}{signal}.
This primarily causes ripples in the objective function surface, leading to troughs and ridges with many local minima and maxima, respectively, as shown in \Subfigref{costsurface}{signal}.
Major shifts in the arrival time of the wave packets lead to overlaps between the different wave packets of the envelopes, see \Subfigref{objectivefunction}{envelope}.
This causes broader valleys and hills in the objective function surface with a reduced number of local minima and maxima, as shown in \Subfigref{costsurface}{envelope}.
The transformation in \Eqref{difference:phase:safe} further diminishes the effect of this overlap by creating an almost monotonically decreasing trend (neglecting $\gamma_k$), where the degrees of freedom of the curve are primarily associated with small changes in the slope.
The information provided by the signals is concentrated in the low-frequency range due to the nature of the autocorrelation.
Both the autocorrelation and stability terms contribute to this regularization.
A direct comparison of the unwrapped phases of the original signals results in sharp discontinuities in the objective function.
Hence, we hypothesize that this mixture of Fourier coefficients in the autocorrelation relaxes and untangles the phases, although we do not have a concrete proof.
All other modifications involving the envelope and analytic signal, absolute values, or phases do not result in broad convexification as in the proposed method.
In this sense, \Eqref{difference:phase:safe} appears to be unique.

\section{Statistical analysis of optimizers}
\label{sec:analysis}

We demonstrate the effectiveness of our proposed methods based on virtual measurements simulated with isotropic material symmetry.
In \Secref{analysis:materialbounds}, an overview of the expected material parameters for the polymers of interest is presented, which serves as a basis for these virtual measurements.
In \Secref{analysis:isotropic}, optimizations for various parameter combinations are performed, and the proposed optimization method is compared with a state-of-the-art optimization method.

\subsection{Material parameter bounds}
\label{sec:analysis:materialbounds}

This study is primarily focused on evaluating the effectiveness of local optimization methods for material parameter estimation.
To properly compare and assess the behavior of these methods numerically, we use virtual measurements, i.e., simulations, as a basis.
Nevertheless, to ensure reproducibility with the real experimental setup and polymer specimens, we carefully restrict the search space to sensible bounds provided in literature and industry for the chosen materials.
This review is performed in part as a preparation for the statistical assessment of the measurement setup and proposed method for transverse isotropic material symmetry.
As such, all parameters required for this extension (including Young's modulus, Poisson's ratio, and shear modulus) are reviewed and presented here.

A thorough investigation has been conducted to identify the typical ranges for density $\rho$, Young's modulus $E$, Poisson's ratio $\nu$, and shear modulus $G$ of the materials of interest: PEEK, PA6, and PP.
In this investigation, we consider already established material parameter bounds provided by manufacturers~\cite{syensqo,kunststoffedirekt,loewen,ensinger,heutecomp,liedtke,tmk,wilhelmherm,chemie,kbkunststoff,amslerfrey,erwintelle,sahlberg,hecksevdic,noltewerk,designerdata,fietz,zellmetall,professionalplastics,polymehr,sonelastic,bpf,ineos,plansee,azo,protoxyz,koenig,pollen,gehr,directplastics,steelplastcc,broncesval} as well as academic sources~\cite{Callister2018}.
We express our gratitude to aggregators such as MatWeb~\cite{Matweb} and Ansys\textsuperscript{\tiny\textregistered}~Granta~\cite{AnsysGranta} for their support.
Owing to licensing issues, the original data cannot be presented in this publication.
Hence, we refer only to the derived quantities based on the reviewed data.

The parameter ranges are limited to typical procedures used for material parameter determination, which involve static or quasi-static experimental setups using standard procedures.
This is somewhat in direct opposition to the intended goal of identifying the parameters in the high-frequency regime.
We argue that the behavior of the optimization methods assessed in this study can be transferred to high-frequency parameter estimation.
We intend to leverage the established knowledge to demonstrate the robustness and effectiveness of our optimization method, with the expectation that these properties are also applicable to real measurements.

Material parameter determination in industrial settings is not only limited to quasi-static experimental setups but also to the scope of material symmetries.
Typically, only the Young's modulus, Poisson's ratio, and/or shear modulus are provided without any further declaration.
As such, we assume that the materials are isotropic in all cases.

\begin{figure}[H]%
	\centering%
	\includegraphics[width=0.98\linewidth]{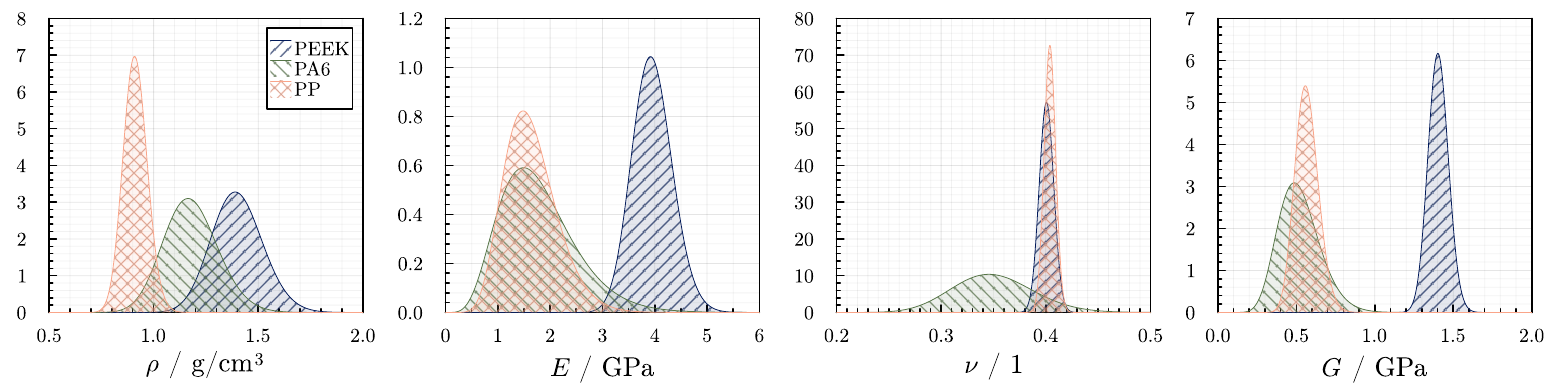}%
	\captionsetup{width=0.\linewidth}
	\caption%
	{
		Gamma distributions of density~$\rho$, Young's modulus~$E$, Poisson's ratio~$\nu$, and shear modulus~$G$ for PEEK, PA6, and PP.
		The numerical values for the shape and scale factors, as well as the expected values and standard deviations for the distributions, are provided in Table~\ref{tbl:parameter:distribution:gamma:values}.
	}%
	\label{fig:parameter:distributions:gamma}%
\end{figure}

We combine the data points of the material parameters by fitting a distribution.
Because material parameters are typically not allowed to be negative, we choose the gamma distribution as the basis\footnote{This does not prohibit a random draw of {$\nu > 0.5$}; however, such an event is extremely unlikely, with a probability of \SI{0.026}{\percent} for the estimated distributions. For all practical purposes, none of the used LHSs include a unit sample (before scaling) of {$> 0.99974$}.}.
The gamma distribution automatically excludes the negative range and is defined as
\begin{equation}\label{eq:distribution:gamma}
	\operatorname{PDF}\subrm{Gamma}(x) = \frac{x^{\alpha-1}}{\Gamma(\alpha) \, \theta^\alpha} \, \exp(-\frac{x}{\theta})
\end{equation}
with shape factor $\alpha$ and scale factor $\theta$.
Here, $\operatorname{\Gamma}(\cdot)$ denotes the gamma function.
Because we include both individual sample points and entire ranges, but we do not know the actual distributions associated with these ranges, we uniformly\footnote{We assume that a uniform distribution is the least biased choice of distribution.} draw 100 samples from them and use such discrete sets of samples to fit the gamma distributions for $\rho$, $E$, $\nu$, and $G$.
We apply a Monte Carlo simulation to improve our inference of the shape and scale factors under the given assumptions.
We repeat the aforementioned process 1000 times and take the averages of $\alpha$ and $\theta$ to acquire their most likely estimates.
We do not include any further statistical analysis of the data but rather only try to confidently identify the most likely values for $\alpha$ and $\theta$ given the data available.
See \Figref{parameter:distributions:gamma} and \Tblref{parameter:distribution:gamma:values} for an overview of the fitted distributions and numerical values, respectively.

\begin{table}[H]%
	\scriptsize
	\centering
	\captionsetup{width=0.7\linewidth, format=plain, labelsep=newline, singlelinecheck=false}
	\caption{
		Values of the shape and scale factors that define the gamma distributions that fit the data for density~$\rho$, Young's modulus~$E$, Poisson's ratio~$\nu$, and shear modulus~$G$ of PEEK, PA6, and PP.
		The expected values and standard deviations derived from the gamma distributions are also presented.
		All values listed here are rounded to five significant digits.
		The respective distributions are shown in \Figref{parameter:distributions:gamma}.
	}\label{tbl:parameter:distribution:gamma:values}
	\begin{tabularx}{0.7\linewidth}{ c l *{4}{S[table-format=1.5e1,table-column-width=20mm]} L{0mm} }
		\toprule
		& & {$\rho \,/\, \si{\gram\per\centi\meter\cubed}$} & {$E \,/\, \si{\giga\pascal}$} & {$\nu \,/\, 1$} & {$G \,/\, \si{\giga\pascal}$} &
		\newrow\cline{2-6}
		\parbox[t]{2mm}{\multirow{4}{*}{\rotatebox[origin=c]{90}{\hspace{-3mm}\textbf{PEEK}}} } &
		  shape factor [-]   & 1.3145e2  & 1.063e2   & 3.2965e3  & 4.7092e2  &
		\newrow
		& scale factor [-]   & 1.0653e-2 & 3.7214e-2 & 1.2158e-4 & 2.9832e-3 &
		\newrow
		& expected value     & 1.4003    & 3.9559    & 0.40079 & 1.4049    &
		\newrow
		& standard deviation & 0.12213 & 0.38368 & 6.9805e-3 & 6.4739e-2 &
		\newrow\cline{2-6}
		\parbox[t]{2mm}{\multirow{4}{*}{\rotatebox[origin=c]{90}{\hspace{-3mm}\textbf{PA6}}}} &
		  shape factor [-]   & 8.3079e1  & 6.0458    & 8.1998e1  & 1.5379e1  &
		\newrow
		& scale factor [-]   & 1.4188e-2 & 0.29571 & 4.268e-3  & 3.3895e-2 &
		\newrow
		& expected value     & 1.1787    & 1.7878    & 0.34997 & 0.52127 &
		\newrow
		& standard deviation & 0.12932 & 0.72711 & 3.8648e-2 & 0.13292 &
		\newrow\cline{2-6}
		\parbox[t]{2mm}{\multirow{4}{*}{\rotatebox[origin=c]{90}{\hspace{-3mm}\textbf{PP}}}} &
		  shape factor [-]   & 2.5313e2  & 1.0516e1  & 5.4154e3  & 5.8031e1  &
		\newrow
		& scale factor [-]   & 3.605e-3  & 0.15586 & 7.46e-5   & 9.7743e-3 &
		\newrow
		& expected value     & 0.91252 & 1.6391    & 0.40399 & 0.56721 &
		\newrow
		& standard deviation & 5.7355e-2 & 0.50544 & 5.4898e-3 & 7.4459e-2 &
		\newrow\bottomrule
	\end{tabularx}
\end{table}

\begin{figure}[H]%
	\centering%
	\begin{subcaptiongroup}%
		\phantomcaption\label{subfig:samples:lhs}%
		\phantomcaption\label{subfig:samples:marginal}%
	\end{subcaptiongroup}%
	\includegraphics[width=0.65\linewidth]{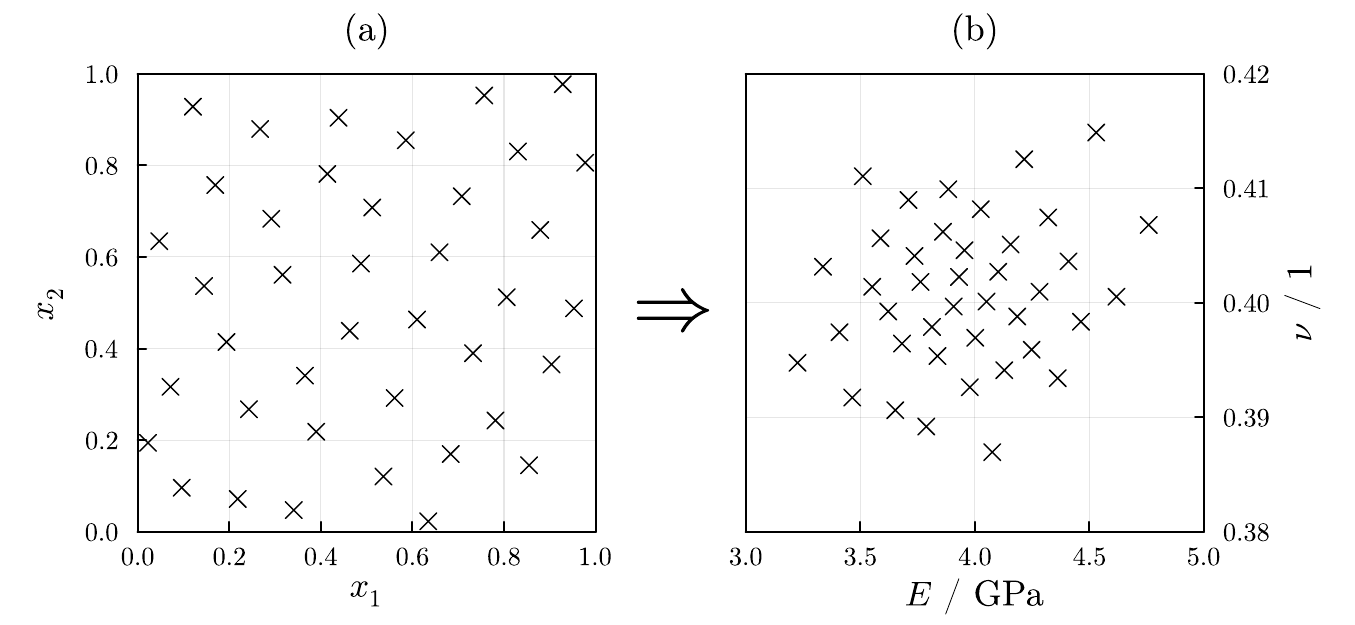}%
	\captionsetup{width=0.65\linewidth}
	\caption%
	{
		Optimized Latin hypercube sampling \subfigref{samples}{lhs} and application of marginal distributions of $E$ and $\nu$ \subfigref{samples}{marginal} for PEEK.
	}%
	\label{fig:samples}%
\end{figure}

The material parameter samples are then drawn from their respective search spaces via an optimized Latin hypercube sampling (LHS)~\cite{Urquhart2020,Bates2012} and subsequently rescaled nonlinearly using the inverse cumulative distribution functions of the marginal gamma distributions.
\Figref*{samples} shows an exemplary draw of $E$ and $\nu$ for PEEK.
The uniformity of the LHS is sacrificed to sample areas of interest more densely.
That is, more likely combinations of parameters are preferred.
These samples are considered ground truths and are used as a basis for simulating fictitious measurements that serve as references for optimization.

Given that the true material parameters are known for these artificial measurements, two different error estimators are provided.
The relative error in the parameter space $\arr{x} \in \reals^m$ is defined via an element-wise relative Manhattan norm, and the relative error in the signal space $\arr{y} \in \reals^n$ is defined as a relative Euclidean norm
\begin{align}
	\operatorname{relative}_1(\arr{x})
	&=
	\norm{\arr{1} - \diag(\hat{\arr{x}})\inverse\arr{x}}_1
	\punct,
	&
	\operatorname{relative}_2(\arr{y})
	&=
	\frac{\norm{\hat{\arr{y}} - \arr{y}}_2}{\norm{\hat{\arr{y}}}_2}
	\punct.
\end{align}
The Manhattan norm is used in conjunction with prenormalization of the parameters to simplify the presentation of the combined optimization convergence behavior by removing the skew due to the difference in the order of magnitude of the parameters.
The Euclidean norm is chosen for the signals because it best aligns with the least-squares objective function.

\subsection{Comparison of optimizers for isotropic material symmetry}
\label{sec:analysis:isotropic}

\begin{figure}[H]%
	\centering%
	\begin{subcaptiongroup}%
		\phantomcaption\label{subfig:optimization:convergence:isotropic:x:peek}%
		\phantomcaption\label{subfig:optimization:convergence:isotropic:x:pa6}%
		\phantomcaption\label{subfig:optimization:convergence:isotropic:x:pp}%
		\phantomcaption\label{subfig:optimization:convergence:isotropic:y:peek}%
		\phantomcaption\label{subfig:optimization:convergence:isotropic:y:pa6}%
		\phantomcaption\label{subfig:optimization:convergence:isotropic:y:pp}%
	\end{subcaptiongroup}%
	\includegraphics[width=0.9\linewidth]{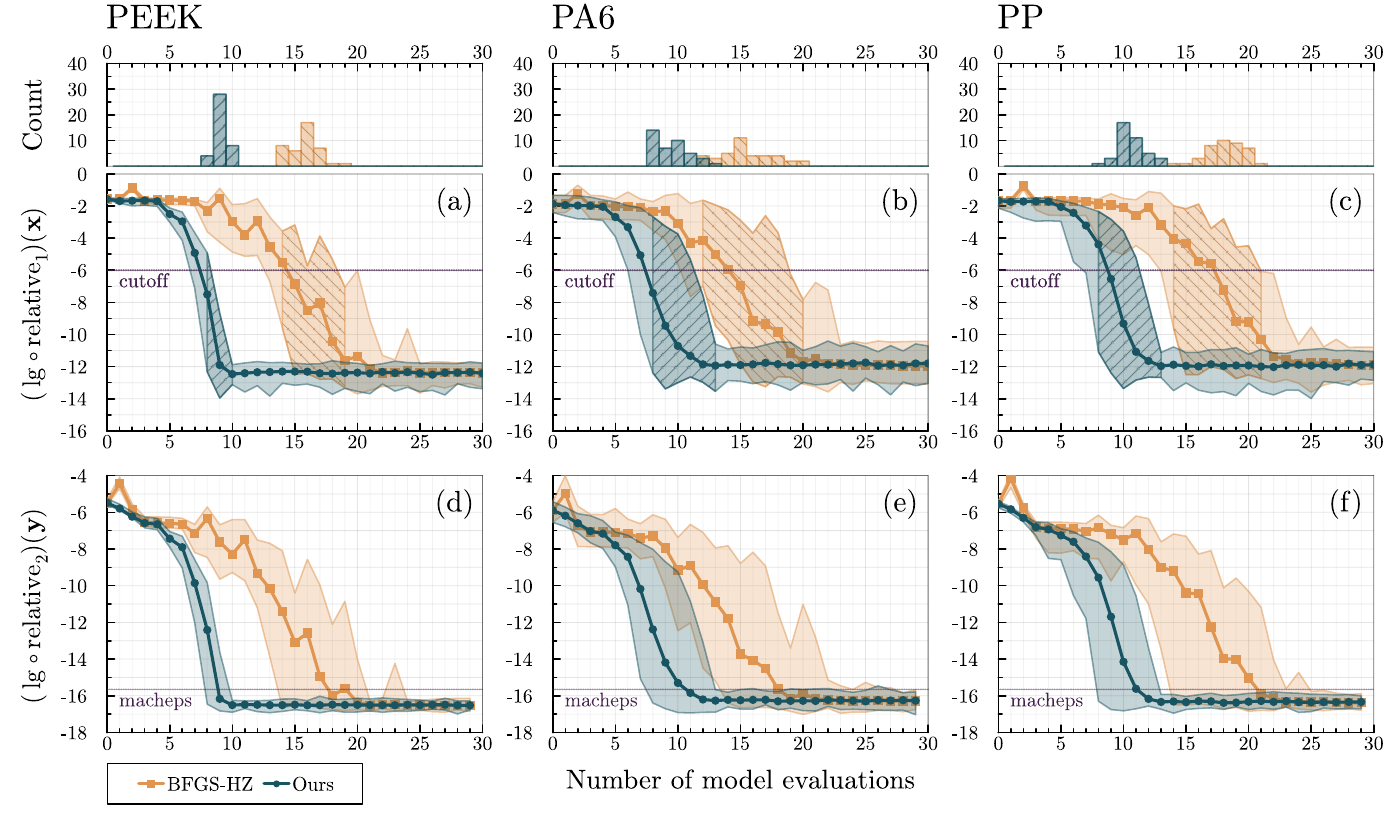}%
	\captionsetup{width=\linewidth}
	\caption%
	{
		Comparison of the BFGS optimizer with the Hager-Zhang line search and our proposed methodology.
		Histograms count the number of optimizations at a minimum number of iterations for which the error in the parameters drops below a cutoff of $10^{-6}$.
		The hull depicts the minimum and maximum errors for 40 samples, and the solid lines indicate the logarithmic mean of all optimizations of a given class.
		The convergence diagrams for the logarithmic relative error between the parameters and measurements include all evaluations of the forward model, i.e., errors of the implicit line search of BFGS are included.
		(This explains the uncharacteristic zig-zag behavior in the convergence, as typically only a drop in the objective function is allowed for BFGS.)
		In general, our method outperforms BFGS, and a sufficiently small error is achieved for all materials in fewer than 15 iterations.
		Given isotropy, both optimization methods succeed within at most 30 iterations.
	}%
	\label{fig:optimization:convergence:isotropic}%
\end{figure}

Material parameter estimation with assumed isotropic material symmetry is the most basic inverse problem that can be solved in this setting.
Provided that the transformation of signals in \Secref{objectivefunction:novel} is used as a basis for the least-squares objective, we demonstrate the convexification of the objective function surface in \Figref{costsurface}.
Hence, the optimization for the isotropic case, given the material bounds provided, should be well-posed.
Therefore, we are confident in using this example to demonstrate the improved performance of our optimization method compared with the state-of-the-art gradient-informed optimization method, the BFGS algorithm. We do not use L-BFGS because only two parameters are optimized.
The precise algorithm deployed is BFGS with a Hager-Zhang line search, as provided by the optimization library Optim.jl~(v1.9.4)~\cite{Mogensen2018}.
We compare the two optimization methods, as shown in \Figref{optimization:convergence:isotropic}.
All initial estimates are generated using the methodology presented in~\cite{Dreiling2024b}.

The BFGS algorithm typically requires interim line searches between Hessian-informed optimization steps.
These line searches require the evaluation of the forward model, and hence, significantly contribute to the total elapsed time used for optimization.
We assume that the forward model is always the bottleneck in terms of compute time; thus, the reduction in the number of evaluations is the driving motivation for rapid convergence.
For this reason, we include these intermediate line search steps and count the total number of model evaluations.
This explains the not strictly monotonic decrease in the optimization objective for BFGS in \Figref{optimization:convergence:isotropic}.
We consider this a more consistent comparison of optimization procedures, as the number of line searches is usually not included.
We believe that the omission of these line search steps is an oversight because the line search significantly contributes to the total number of model evaluations and, ultimately, the total compute time.

To statistically evaluate and compare the behaviors of the optimization methods, a total of $60$ simulated measurements are produced and serve as virtual measurement references.
Optimizations are deemed successful if the relative deviation in the material parameters drops below a cutoff of $10^{-6}$, and the associated minimum number of iterations is used for comparing the optimizers.
On average, our optimization method outperforms BFGS and requires fewer total iterations to converge to the correct solution.
This effect is most pronounced for PEEK, as shown in \Subfigref{optimization:convergence:isotropic}{x:peek}~and~\subfigref{optimization:convergence:isotropic}{y:peek}.
Minor overlaps in the distributions of the iteration counts exist for PA6~(\Subfigref{optimization:convergence:isotropic}{x:pa6}~and~\subfigref{optimization:convergence:isotropic}{y:pa6}) and PP~(\Subfigref{optimization:convergence:isotropic}{x:pp}~and~\subfigref{optimization:convergence:isotropic}{y:pp}).
However, these differences are negligible.

The overall behavior of our optimizer shows rapid convergence.
In addition, we can demonstrate that the least-squares objective is a good measure of fit, and its convergence behavior correlates well with the underlying deviations from the true material parameter combinations.

\section{Conclusion}
\label{sec:conclusion}

The contributions of this study are twofold:
First, we introduced a novel automatic step adaption for the standard gradient descent and Levenberg–Marquardt methods.
In a qualitative comparison, we were able to show that our choice of optimization method leads to increased performance w.r.t.\ the convergence rate of optimization for this particular problem setup by reducing the number of optimization steps.

Second, we provide a novel transformation based on unwrapped phases for stable signal comparisons for material parameter determination.
Notably, for the isotropic case, our transformation yields convexified objective function surfaces for different references and allows stable and rapid optimization.
In principle, most first-order optimization methods should rapidly converge to the correct solution.

However, it is unclear whether the proposed transformation yields equally stable optimizations in the transverse isotropic case with damping.
This more complicated material model may reintroduce such nonlinearities in the objective function, which can destabilize first-order optimization methods.
To this end, we preemptively propose a mixture of gradient descent and Gauss-Newton methods in the style of Levenberg–Marquardt to utilize the excellent optimization behavior of gradient descent far from a minimizer and the superior convergence rate of the Gauss-Newton method close to a minimizer.

Future research topics include the analysis of the optimization behavior of transverse isotropic material symmetry with damping, an assessment of the influence of the experimental setup, specifically symmetry-breaking configurations, and experimental validation of the measurement setup and optimization method.

\begin{acknowledgements}
	\item Funded by the Deutsche Forschungsgemeinschaft (DFG, German Research Foundation) — 409779252.
	\item The authors are grateful for the compute time provided by the Paderborn Center for Parallel Computing (PC2).
	\item The Julia packages Plots.jl~\cite{Breloff2025} and Makie.jl~\cite{Danisch2021} were used to generate some of the figures presented in this paper.
\end{acknowledgements}

\section*{\color{header}References}
\begin{multicols}{2}\color{header}\printbibliography[heading=none]\end{multicols}

\end{document}